\documentclass[prx,floatfix,twocolumn]{revtex4-2}
\usepackage{ORI_Group_style}
\usepackage{mathtools}
\usepackage{tabularx}
\usepackage{natbib}
\usepackage{booktabs}
\usepackage{xcolor}
\newcommand{\uv}{\mathbf{e}}
\newcommand{\ForceVecOp}{ \hat{\boldsymbol{\mathcal{F}}}}
\newcommand{\ForceOp}{ \hat{\mathcal{F}}}
\newcommand{\LambdaNameCapitalized}{Recoil localization parameter }
\newcommand{\LambdaName}{recoil localization parameter }

\hypersetup{citecolor = blue, colorlinks = true}

\begin{document}

\title{Quantum Theory of Light Interaction with a Lorenz-Mie Particle: \\ Optical Detection and Three-Dimensional Ground-State Cooling}

\author{Patrick~Maurer}
\affiliation{Institute for Quantum Optics and Quantum Information of the Austrian Academy of Sciences, A-6020 Innsbruck, Austria.}
\affiliation{Institute for Theoretical Physics, University of Innsbruck, A-6020 Innsbruck, Austria.}

\author{Carlos~Gonzalez-Ballestero}
\affiliation{Institute for Quantum Optics and Quantum Information of the Austrian Academy of Sciences, A-6020 Innsbruck, Austria.}
\affiliation{Institute for Theoretical Physics, University of Innsbruck, A-6020 Innsbruck, Austria.}

\author{Oriol~Romero-Isart}
\affiliation{Institute for Quantum Optics and Quantum Information of the Austrian Academy of Sciences, A-6020 Innsbruck, Austria.}
\affiliation{Institute for Theoretical Physics, University of Innsbruck, A-6020 Innsbruck, Austria.}

\begin{abstract}

We theoretically analyze the motional quantum dynamics of a levitated dielectric sphere interacting with the quantum electromagnetic field beyond the point-dipole approximation. To this end, we derive a Hamiltonian describing the fundamental coupling between photons and center-of-mass phonons, including Stokes and anti-Stokes processes, and the coupling rates for a dielectric sphere of arbitrary refractive index and size. We then derive the laser recoil heating rates and the information radiation patterns (the angular distribution of the scattered light that carries information about the center-of-mass motion) and show how to evaluate them efficiently in the presence of a focused laser beam, either in a running or a standing-wave configuration. This information is crucial to implement active feedback cooling of optically levitated dielectric spheres beyond the point-dipole approximation. Our results predict several experimentally feasible configurations and parameter regimes where optical detection and active feedback can simultaneously cool to the ground state the three-dimensional center-of-mass motion of dielectric spheres in the micrometer regime. Scaling up the mass of the dielectric particles that can be cooled to the center-of-mass ground state is not only relevant for testing quantum mechanics at large scales but also for current experimental efforts that search for new physics (e.g. dark matter) using optically levitated sensors.

\end{abstract}

\maketitle

\section{Introduction}

The interaction of light with a dielectric sphere is not only a paradigm in optical physics~\cite{Lorenz1890,Mie1908, AshkinPRL1970,Ashkin1971,AshkinAPL1976,Ashkin2006} but a topic of current research in the field of levitated optomechanics~\cite{GonzalezBallestero2021,MillenRepProgPhys2020}. Particularly relevant are recent experiments achieving ground-state cooling of the center-of-mass motion  of an optically trapped dielectric nanosphere using either the coherent coupling to an optical resonator~\cite{Delic2020, Ranfagni2022, Piotrowski2022} or shot-noise limited optical detection and active feedback~\cite{Magrini2021,Tebbenjohanns2021, Kamba2022}. These experiments have been limited to ground-state cooling of one and two center-of-mass degrees of freedom of sub-wavelength silica nanoparticles, that is spheres with a radius $R$ of around $100$~nm (mass of $10^9$ atomic mass units), thus much smaller than the optical wavelength $\lambda_0$ of the laser light.  At these scales, where $R/\lambda_0 \ll 1$, the electrodynamical response of the dielectric particle can be approximated by a point dipole. Within this point-dipole approximation, the understanding of how light interacts with the motion of the nanoparticle (e.g. optomechanical coupling rates, laser recoil heating rates, information radiation patterns)~\cite{RomeroIsartNJP2010,ChangPNAS2010,Romero-Isart2011,Rodenburg2016,Tebbenjohanns2019,Ballestero2019,Rudolph2021,Toros2021} has been key to predict, optimize, and understand ground-state cooling. This theory is, however, rather limited and can not be easily extended to dielectric particles for which the point-dipole approximation is not valid~\cite{PflanzerPRA2012,Seberson2020}. Therefore, it has hitherto remained unclear whether larger dielectric spheres with a radius comparable or larger than the laser light wavelength can be cooled to the ground state, and if so, how to achieve it. 

In this article we develop a quantum electrodynamical theory of the interaction between the electromagnetic field and a dielectric sphere of arbitrary refractive index and size. One of the main results of this theory is to show that simultaneous three-dimensional center-of-mass ground-state cooling of a dielectric sphere with a size ranging from few hundreds of nanometers to several micrometers is possible using optical setups that are currently implemented in laboratories~\cite{Tebbenjohanns2019CD,Conangla2019,Magrini2021,Ciampini2021,Tebbenjohanns2021,Kamba2021,Kamba2022}: shot-noise limited optical detection of either forward or backward scattered light of a focused laser beam in a running or a standing wave configuration. Our results are thus timely and open the door to bringing dielectric spheres of masses ranging from $10^9$ to $10^{14}$ atomic mass units into the quantum regime using laser light at room temperature~\cite{Ashkin1977,Li2011,BlakemorePRA2019,MonteiroPRA2020,Arita2022}. 
 Scaling up the mass of the dielectric spheres that can be laser cooled to the ground state has applications in the search for new physics (e.g. dark matter~\cite{RademacherAOT2020,KawasakiRSI2020,MonteiroPRL2020,AfekPRD2021,BlakemorePRD2021,MooreQST2021,AfekPRL2022,PrielSciAdv2022,Carney2022}). In addition, in combination with non-optical potentials (e.g. electrostatic~\cite{Millen2015,Conangla2020,Lorenzo2021}), one could consider to delocalize their center-of-mass to scales orders of magnitude larger than their zero-point motion~\cite{Weiss2021, Roda-Llordes_in_preparation}, thereby testing quantum mechanics at unprecedented mass scales~\cite{Romero-Isart2011LQ,Romero-Isart2011QS, Bateman2014,Wan2016,Neumeier2022}, comparable to current and planned efforts with superconducting microspheres~\cite{Romero-IsartPRL2012,Romero-Isart2017,Pino2018,Hofer2022,Latorre2022}.

More specifically, our article contains five key results in the field of levitated optomechanics~\cite{GonzalezBallestero2021,MillenRepProgPhys2020} and is organized as follows. (i)~We analytically derive the optomechanical coupling rates for arbitrary dielectric spheres (Sec.~\ref{sec:fundamentalHamiltonian}). Crucially, this is enabled by our previous work~\cite{Maurer2021}, where using the techniques of Ref.~\cite{Glauber1991} we quantize the electromagnetic field in the presence of a non-moving sphere in terms of normalized scattering eigenmodes. (ii)~Taking the small-particle limit, we provide a rigorous justification for the phenomenological light-matter Hamiltonian, based on the point-dipole approximation, used thus far in levitated optomechanics~\cite{RomeroIsartNJP2010,ChangPNAS2010,Romero-Isart2011,Rodenburg2016,Tebbenjohanns2019,Ballestero2019,Rudolph2021, Toros2021}~(Sec.~\ref{sec:smallParticleLimit}). (iii)~We derive expressions for the recoil heating rates and the information radiation patterns~\cite{Tebbenjohanns2019,Magrini2021,Tebbenjohanns2021}, two core ingredients in levitated optomechanics that we define here in the context of transition amplitudes of Stokes and anti-Stokes processes, for arbitrary particle sizes, refractive indices, and for an arbitrary configuration of laser fields~(Sec.~\ref{sec:recoilHeatingRates}). Furthermore, we simplify their expressions analytically which enables us to evaluate them efficiently. (iv)~We compute recoil heating rates and information radiation patterns in two configurations of direct experimental interest~(Sec.~\ref{sec:case_study}), namely a single focused Gaussian beam and two counter-propagating focused beams in a standing wave configuration (Sec.~\ref{sec:runningWave} and Sec.~\ref{sec:standingWave}). We show results beyond the point-dipole approximation and beyond the plane-wave approximation for the illuminating beams, and characterize their dependence on particle radius and numerical aperture of the focusing lens. (v)~We show how the results of (iv) predict a broad parameter regime where one, two, and even three-dimensional center-of-mass ground-state cooling is possible for dielectric spheres ranging from few hundreds of nanometers to few micrometers~(Sec.~\ref{sec:case_study}). Finally, in the conclusions (Sec.~\ref{Sec:Conclusions}) we argue that the methods developed in this work could directly be extended to other particle shapes and degrees of freedom~\cite{Arita2013,Kuhn2015,Hoang2016,Kuhn2017,Kuhn2017/2,Monteiro2018,Rashid2018,Ahn2018,Ahn2020,Laan2020,Pontin2022}, thus providing a complete theoretical toolbox to describe the interaction of light with levitated dielectric objects in the quantum regime.

\section{Fundamental Hamiltonian \& photon-phonon coupling rates} \label{sec:fundamentalHamiltonian}

Let us consider a dielectric sphere of radius $R$ and mass $M$ whose equilibrium center-of-mass position is by definition at the origin of coordinates.  The center-of-mass position fluctuations around the equilibrium position are described by the quantum mechanical position and momentum operators $\hat \rr$  and $\hat \pp$, respectively. They fulfil the commutation relations $\com{\hat r_\mu}{ \hat p_\nu} = \im \hbar \delta_{\mu \nu} $, where the indices $\mu,\nu$ label the coordinate axes $x,y,z$, $\delta_{\mu \nu}$ is the Kronecker delta, and $\hbar$ the reduced Planck's constant. The sphere is assumed to be in vacuum, interacting only with electromagnetic fields.  We assume that the non-fluctuating part of the electromagnetic field (e.g. optical tweezer) generates a conservative potential for the center-of-mass motion given by $V(\rr)$. We model the electromagnetic response of the dielectric sphere as homogeneous, isotropic, and lossless. We assume that the dielectric sphere interacts significantly only with electromagnetic field modes in a sufficiently narrow frequency window, so that its electromagnetic response can be described by a single scalar and real relative permittivity $\epsilon$. Assuming sufficiently small center-of-mass displacements, that is $\Delta r_\mu \equiv \avg{\hat r_\mu^2- \avg{\hat r_\mu}^2}^{1/2}$ smaller than any relevant length scale associated to the electromagnetic fields, the total Hamiltonian describing the interaction between the center-of-mass motion and the electromagnetic field can be expressed as
\be \label{eq:QEDHamiltonian}
\hat H = \frac{\hat \pp^2}{2M} + V(\hat \rr) + \hat{H}_\text{em} -\ForceVecOp \cdot \hat \rr.
\ee
Here $\hat{H}_\text{em}$ describes the free dynamics of the electromagnetic field in the presence of a fixed dielectric sphere at the origin of coordinates, the derivation and thorough discussion of which is the focus of our previous work~\cite{Maurer2021}. The last term, which is the focus of this article, describes the interaction of the electromagnetic field with the center-of-mass position. Note that the small-displacement assumption manifests in an interaction term that is linear in the position operators and depends on the specific form of the operator $\ForceVecOp$, which hereafter we call {\em radiation pressure operator}. As we show below $\ForceVecOp$ depends on the electromagnetic field degrees of freedom, is nonlinear with the electromagnetic fields, and does not commute with $ \hat{H}_\text{em}$, namely $ \coms{\hat{H}_\text{em}}{\ForceOp_\mu} \neq 0$ for any $\mu$. 
We remark that by definition of $V(\rr)$, we have that
\be \label{eq:potential_condition}
\grad V (\rr) |_{\rr=0} = \avg{ \ForceVecOp}.
\ee
Here $\avg{ \ForceVecOp}$ represents  the quantum mechanical expected value of $\ForceVecOp$ for a given state of the electromagnetic field and the time average over the relevant timescale of the electromagnetic field. The relevant electromagnetic timescale (e.g. optical time period) is assumed much shorter than the mechanical timescales associated to the motion of the dielectric sphere. In the context of discussing center-of-mass cooling, we will consider the potential $V(\rr)$ to be a standard three-dimensional anisotropic harmonic potential with harmonic frequencies $\Omega_\mu$, namely
\be
V(\hat \rr) = \frac{M}{2}  \sum_\mu \Omega^2_\mu \hat r_\mu^2.
\ee
We write position and momentum operators in terms of bosonic creation $\bdop_\mu$ and annihilation $\bop_\mu$ operators, namely $\hat r_\mu = r_{0\mu} (\bdop_\mu + \bop_\mu)$, and $\hat{p}_\mu = \im M \Omega_\mu r_{0\mu}(\bdop_\mu - \bop_\mu)$, where $r_{0\mu} \equiv [\hbar/(2 M \Omega_\mu)]^{1/2}$ is the zero-point motion and $\coms{\bop_\mu}{ \bdop_\nu} = \delta_{\mu \nu}$.

Let us now discuss the specific form of the radiation pressure operator $\ForceVecOp$. Using \eqnref{eq:QEDHamiltonian}, one can show that the Heisenberg equation of motion for the $\mu$ component of the center-of-mass position operator of the dielectric sphere is given by
\begin{equation} \label{eq:eom_RP}
	\frac{\text{d}^2\hat{r}_\mu}{\text{d}t^2}+ \Omega^2_\mu \hat{r}_\mu =\frac{\ForceOp_\mu}{M}.
\end{equation}
We obtain an expression for the radiation pressure operator by deriving the equivalent classical equation of motion according to electrodynamics in the presence of a dielectric object, promoting the dynamical variables (position, momentum, and electromagnetic fields) to quantum operators, and comparing the resulting equation with \eqnref{eq:eom_RP}. The right-hand side of such equation is the radiation pressure exerted by the self-consistent electromagnetic fields, which in electrodynamics can be expressed by a surface integral of the Maxwell stress tensor in the far field of the dielectric sphere~\cite{Jackson1999,novotny_hecht_2012}. Based on this, we define the vector component of the radiation pressure operator $\ForceOp_\mu$, written in terms of the electric $\hat{\EE}(\rr)$ and magnetic $\hat{\BB}(\rr)$ field operators, as
\begin{equation} \label{eq:definition_F}
    \ForceOp_\mu \equiv -\frac{\epsilon_0}{2} \lim_{r\to\infty}r^2 \int \text{d}\Omega (\uv_r\cdot \uv_\mu) [\hat{\EE}^2(\rr)+c^2\hat{\BB}^2(\rr)].
\end{equation}
Here $r = |\rr|$, $\text{d}\Omega $ is the surface element in spherical coordinates, $\uv_r$  the radial unit vector,  $\uv_\mu$  the $
\mu$-axis unit vector, and $\epsilon_0$ and $c$ the permittivity and speed of light in vacuum, respectively.  The expression in \eqnref{eq:definition_F} fully determines the fundamental Hamiltonian in \eqnref{eq:QEDHamiltonian} describing the interaction of light with the center-of-mass motion of a dielectric sphere of arbitrary refractive index and size in the small-displacement regime.

Let us now expand the electric and magnetic field operators in terms of the normalized eigenmodes for a fixed sphere at the origin of coordinates, namely
\begin{align}
\label{eq:electric_operator_eigenmodes}
\hat{\mathbf{E}}(\rr)&=\im\sum_{\kappa}\sqrt{\frac{\hbar \omega_\kappa}{2\epsilon_0}}\spare{\mathbf{F}_\kappa(\rr)\hat{a}_\kappa-\text{H.c.}},\\
\label{eq:magnetic_operator_eigenmodes}
\hat{\mathbf{B}}(\rr)&=\sum_{\kappa}\sqrt{\frac{\hbar}{2\epsilon_0\omega_\kappa}}\spare{\nabla\times\mathbf{F}_\kappa(\rr)\hat{a}_\kappa+\text{H.c.}}.
\end{align}
Here, we denote by $\mathbf{F}_\kappa(\rr)$ the normalized scattering eigenmodes in the presence of a non-moving dielectric sphere, whose calculation and quantization for arbitrary particle sizes and refractive indices was the object of our previous work~\cite{Maurer2021}. 
The normalized scattering eigenmodes are defined by the multi-index $\kappa = (g,\kk)$, where $g =1,2$ is a polarization index with associated polarization unit vector $\uv_{g}$ and $\kk \in \mathbb{R}^3$ is the wave vector. They are composed by a plane-wave contribution with mode index $\kappa$ (i.e. wave vector $\kk$ and transverse polarization vector $\uv_{g}$) and a scattered contribution describing the elastic scattering of the plane-wave component off the dielectric sphere. In essence, they are properly normalized solutions of the Lorenz-Mie problem~\cite{Lorenz1890,Mie1908}, which allow us to fully include all the elastic scattering processes (i.e. the scattering of light by the non-moving dielectric particle) exactly. The mode frequency is given by $\omega_\kappa = c |\kk| $. The associated creation and annihilation operators fulfil the bosonic commutation  rules $\coms{\aop_\kappa}{\adop_{\kappa'}} = \delta_{\kappa \kappa'}$~\footnote{We remark that throughout the article we will use the  notation wherein sums over a multi-index $\kappa$ must be understood as integrals over continuous indices and sums over discrete ones, namely $\sum_\kappa \equiv \int \text{d} \kk \sum_g$. A Kronecker delta of a multi-index must be understood as a product of Dirac deltas for the continuous indices and Kronecker deltas for discrete indices, namely  $\delta_{\kappa \kappa'} \equiv \,\delta(\kk - \kk') \delta_{gg'}$.}. In terms of the normalized eigenmodes, the Hamiltonian $\Hop_\text{em}$ describing the free dynamics of the electromagnetic field for a fixed sphere is diagonal, namely $\Hop_\text{em} = \sum_\kappa \hbar \omega_\kappa \adop_\kappa \aop_\kappa$, for spheres of all refractive indices and sizes~\cite{Maurer2021}. The expansion in terms of normalized scattering eigenmodes is a crucial step to obtain exact analytical expressions of quantities of interest and is in contrast to usual field expansion in terms of e.g. plane waves, for which the Hamiltonian is not diagonal. Such cases require the use of approximative expansions which fail for large enough particles~\cite{PflanzerPRA2012}.

\begin{figure}[htb!]
    \centering
    \includegraphics[width=0.9\columnwidth]{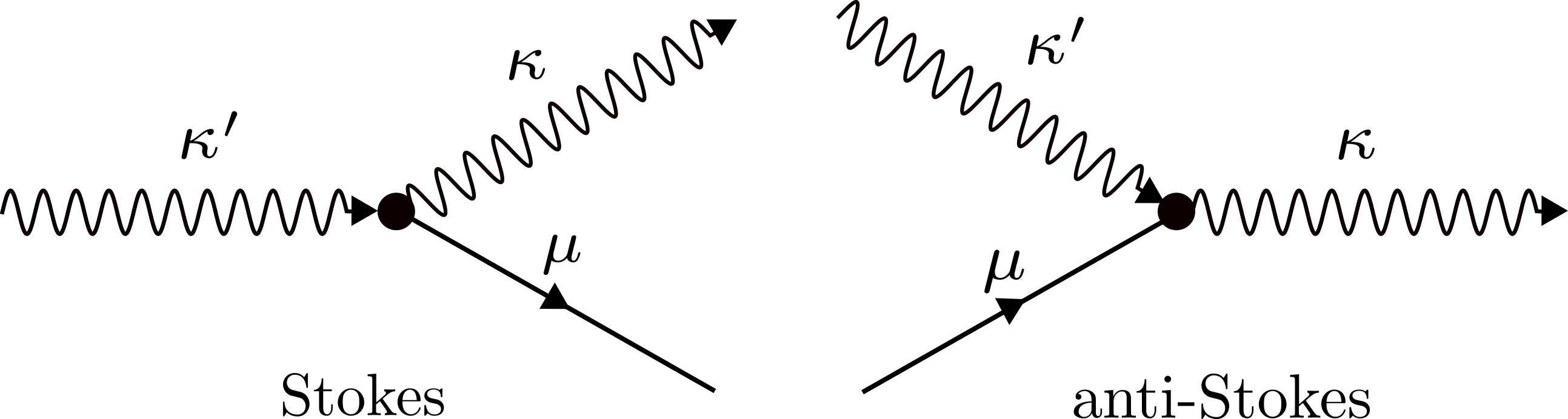}
    \caption{Feynman diagrams of the two fundamental interaction processes as described by the third term in \eqnref{eq:QEDHamiltonianBosons}. Photons in mode $\kappa'$ are inelastically scattered at a coupling rate $g_{\kappa\kappa'\mu}$ into mode $\kappa$ through generation (Stokes) or absorption (anti-Stokes) of a phonon in mode $\mu$.}
    \label{fig1}
\end{figure}

We can now express the Hamiltonian \eqnref{eq:QEDHamiltonian} in terms of bosonic creation and annihilation operators only, resulting in a Hamiltonian that describes the dynamics of photons interacting with center-of-mass phonons. Hereafter, we refer to a photon as the excitation of an eigenmode of the electromagnetic field in the presence of the dielectric sphere at equilibrium. This Hamiltonian is obtained by introducing \eqnref{eq:electric_operator_eigenmodes} and \eqnref{eq:magnetic_operator_eigenmodes} into  \eqnref{eq:definition_F} and performing a rotating-wave approximation that neglects all rapidly oscillating terms. The rotating-wave approximation is valid in the standard weak-coupling regime where the bare photon-phonon interaction coupling rate is smaller than the phonon frequencies~\footnote{Note that the rotating-wave approximation also requires the frequencies of the electromagnetic field to be large compared to the phonon frequencies. This condition is satisfied for the narrow-band and high-frequency optical modes considered in this work.}.  The Hamiltonian  \eqnref{eq:QEDHamiltonian} can then be written as
\begin{multline} \label{eq:QEDHamiltonianBosons}
\Hop= \sum_\mu \hbar\Omega_\mu \bdop_\mu \bop_\mu + \sum_\kappa \hbar\omega_\kappa \adop_\kappa \aop_\kappa \\ + \hbar\sum_{\kappa \kappa' \mu} g_{\kappa \kappa' \mu} \adop_\kappa \aop_{\kappa'} (\bdop_\mu + \bop_\mu).
\end{multline}
The third term describes the two fundamental processes describing the interaction of light and center-of-mass motion as illustrated in \figref{fig1}, that is (i) Stokes processes, in which a photon in mode $\kappa'$ is inelastically scattered into a photon of lower frequency in mode $\kappa$ by generating a phonon in mode $\mu$, and (ii) anti-Stokes processes, in which a photon in mode $\kappa'$ is inelastically scattered into a photon of higher frequency in mode $\kappa$ by absorbing a phonon in mode $\mu$. 

The physics of the photon-phonon interaction is encoded in the coupling rates $g_{\kappa \kappa' \mu}$ of \eqnref{eq:QEDHamiltonianBosons}, which can be written as
\begin{multline} \label{eq:gcouplings}
g_{\kappa\kappa'\mu}= r_{0 \mu} \frac{c\sqrt{kk'}}{2} \lim_{r\to\infty}r^2\int \text{d}\Omega(\uv_r\cdot \uv_\mu) \\ \times
\left[\FF^*_{\kappa}(\rr)\cdot\FF_{\kappa'}(\rr)+\frac{1}{kk'}\nabla\times\FF^*_{\kappa}(\rr)\cdot\nabla\times\FF_{\kappa'}(\rr)\right],
\end{multline}
with $k=|\kk|$. That is, they correspond to a surface integral in the far-field of a function that depends on the normalized scattering eigenmodes~\cite{Maurer2021}. As shown in \appref{App:explicit_coupling}, the far-field expression of the scattering modes can be written in terms of vector spherical harmonics with angular momentum mode indices $l=1,2,\ldots$ and $m = - l, -l+1,\ldots,l$ which can be mapped to the angular distribution of the electromagnetic fields radiated by electric and magnetic multipoles. The angular integrals \eqnref{eq:gcouplings}, containing products of vector spherical harmonics, can be analytically performed using the results of~\cite{Neves2019}. This means that the couplings $g_{\kappa \kappa' \mu}$ for a dielectric sphere of arbitrary refractive index and size can be analytically expressed as a sum over discrete angular momentum and polarization indices that can be efficiently evaluated, see \eqnref{eq:couplingratesANotation}. This is one of the main results of this article. Let us remark that we have explicitly confirmed that the Hamiltonian \eqnref{eq:QEDHamiltonianBosons} with the coupling rates \eqnref{eq:gcouplings} can also be obtained applying perturbation theory to the electromagnetic Hamiltonian, as described in~\cite{Sipe2016,Zoubi2016} in the context of Brillouin scattering in photonic waveguides.

\subsection{Coupling rates in the small-particle limit}\label{sec:smallParticleLimit}

As a consistency check, one can show (details in~\appref{App:explicit_coupling}) that in the small particle limit $\sqrt{\epsilon}k R \rightarrow 0$, the lowest-order term of the photon-phonon coupling rates \eqnref{eq:gcouplings} can be written as
\begin{equation} \label{eq:gcouplings_small_particle}
	g_{\kappa\kappa'\mu}\simeq \frac{\im r_{0\mu} \alpha}{\epsilon_0} \frac{c k k'  }{2 (2\pi)^3}(\uv_{g}^*\cdot \uv_{g'})(\uv_k-\uv_{k'})\cdot \uv_{\mu},
\end{equation}
where $\alpha = 3\epsilon_0 V(\epsilon-1)/(\epsilon+2)$ is the polarizability of the dielectric sphere, $V$ denotes the volume of the sphere, and $\uv_k$ denotes the unit vector parallel to the wave vector $\kk$ . The consistently derived small-particle coupling rates~\eqnref{eq:gcouplings_small_particle} agree with the coupling rates used in the current literature~\cite{RomeroIsartNJP2010,ChangPNAS2010,Romero-Isart2011,PflanzerPRA2012,Rodenburg2016,Ballestero2019,Rudolph2021, Toros2021}, where they are {\em heuristically} derived by making a Taylor expansion of the interacting Hamiltonian $-\alpha\mathbf{\hat{E}}^2(\hat{\rr})/2$ with the electric field operator expanded in plane-wave modes (point-dipole approximation). The latter can be obtained by replacing the scattering normalized modes $\mathbf{F}_\kappa(\rr)$ in \eqnref{eq:electric_operator_eigenmodes} by plane-wave modes $\mathbf{G}_\kappa(\rr) \equiv \exp (\im \kk \cdot \rr) \uv_{g} (2 \pi)^{-3/2}$. We find this agreement an important result from a theoretical point of view, as it provides a sound justification of the interacting Hamiltonian $-\alpha\mathbf{\hat{E}}^2(\hat{\rr})/2$ used to describe the interaction of light with a small dielectric particle in the quantum regime. Furthermore, the comparison of \eqnref{eq:gcouplings_small_particle} with \eqnref{eq:gcouplings} provides a clear route to make predictions that go beyond the point-dipole approximation and that could be observed in current experiments, as we show in the rest of the article.

\subsection{Linearized Hamiltonian in presence of a classical field}\label{sec:linearization}

Let us consider the scenario where some of the electromagnetic field modes are in a coherent state. That is, the quantum state of the electromagnetic field in presence of the nonmoving sphere at the origin is given by
\begin{equation} \label{eq:Eclassical}
\ket{\Psi_\text{cl}}  = \exp \spare{\sum_\kappa (\alpha_\kappa \hat{a}_\kappa^\dagger -\alpha_\kappa^* \hat{a}_\kappa)} \ket{0}   .
\end{equation}
Here $\alpha_\kappa$ is the coherent complex amplitude of the mode $\kappa$, that is $\aop_\kappa \ket{\Psi_\text{cl}}  = \alpha_\kappa \ket{\Psi_\text{cl}} $. The distribution in $\kappa$ of $\alpha_\kappa$ specifies the particular state of the electromagnetic field. The associated classical electromagnetic field to this coherent quantum state is given by
\begin{multline}
\label{eq:classicalField}
\mathbf{E}_\text{cl}(\rr,t) \equiv \bra{\Psi_\text{cl}} \hat{\mathbf{E}}(\rr,t) \ket{\Psi_\text{cl}}  \\ =\im\sum_{\kappa}\sqrt{\frac{\hbar \omega_\kappa}{2\epsilon_0}}\spare{\mathbf{F}_\kappa(\rr)\alpha_\kappa e^{- \im \omega_\kappa t} -\text{c.c.}}.
\end{multline}
 Since in most of the experiments in levitated optomechanics the classical electromagnetic field is generated by a quasi-monochromatic laser light, hereafter we will assume that all the modes found in this coherent state have the same frequency, labelled by $\omega_0$. As it is common in optomechanics, one can now derive the linearized Hamiltonian describing the dynamics of the fluctuations of the electromagnetic field above $\ket{\Psi_\text{cl}}$ caused by their interaction with the center-of-mass motion of the dielectric sphere. In a rotating frame with frequency $\omega_0$, this linearized Hamiltonian is given by
  \begin{multline}\label{eq:linearizedHamiltonian}
 	\hat{H}_\text{lin}=\sum_\kappa \hbar\Delta_\kappa \hat{a}^\dagger_\kappa\hat{a}_\kappa+ \sum_\mu \hbar\Omega_\mu \hat{b}^\dagger_\mu\hat{b}_\mu\\
 + \hbar\sum_{\kappa\mu}(G_{\kappa\mu}\hat{a}_\kappa^\dagger + G^*_{\kappa\mu}\hat{a}_\kappa) (\hat b^\dagger_\mu + \hat b_\mu),
 \end{multline} 
 where we have defined the detuning $\Delta_\kappa\equiv\omega_\kappa-\omega_0$ for each mode and the linearized coupling rate 
 \begin{equation}\label{eq:linearizedCouplingRate}
     G_{\kappa\mu}\equiv \sum_{\kappa'}\alpha_{\kappa'}g_{\kappa\kappa'\mu}.
 \end{equation}
We remark that the potential describing the center-of-mass motion in the presence of the electromagnetic field $\ket{\Psi_\text{cl}}$ is defined so that  \eqnref{eq:potential_condition} is fulfilled, that is, it includes the radiation pressure exerted by the classical electromagnetic field. Indeed, in most cases the classical electromagnetic field $\mathbf{E}_\text{cl}(\rr,t)$ represents the field used for optical trapping.

As we show in the next section, the fundamental coupling rates $g_{\kappa\kappa'\mu}$ (\eqnref{eq:gcouplings}) and the distribution of coherent amplitudes $\alpha_\kappa$ for a given classical electromagnetic field (\eqnref{eq:Eclassical}), are the key ingredients to analyze the  quantum dynamics of a trapped dielectric sphere interacting with light.

\section{Recoil Heating rates \& Information Radiation Patterns}\label{sec:recoilHeatingRates}

As shown in the previous section, the fundamental transitions occurring during the interaction of light with a trapped particle are given by Stokes and anti-Stokes processes, see \figref{fig1}. These are the only processes that change the center-of-mass state of the particle. In the presence of a highly excited electromagnetic field, such as the classical electromagnetc field given by $\ket{\Psi_\text{cl}}$ (\eqnref{eq:Eclassical}), Stokes and anti-Stokes processes are significant, and the overall effective dynamics is described by the linearized Hamiltonian~\eqnref{eq:linearizedHamiltonian}. These effective dynamics yield two important phenomena: (i) dissipative center-of-mass dynamics, namely laser light recoil heating, and (ii) the scattering of light carrying information about the center-of-mass position fluctuations, the angular distribution of which is called information radiation pattern (IRP). 

To better define and evaluate these two quantities, let us consider the transition amplitude for the Stokes and anti-Stokes process in the presence of a classical electromagnetic field. The transition amplitudes associated to these processes can be evaluated as~\cite{Cohen2004/2}
\be\label{eq:asymptotic_TA}
\tau^p_{\kappa \mu}\equiv  \bra{\Psi^p_\text{out}}\hat{U}(T/2,-T/2)\ket{\Psi_\text{in}}.
\ee 
Here the input state is given by $\ket{\Psi_\text{in}}\equiv \ket{n_x,n_y,n_z} \otimes \ket{\Psi_\text{cl}}$, where $\ket{n_x,n_y,n_z}$  ($n_\mu >0$) is a product Fock state for the three center-of-mass degrees of freedom and the quantum state of the electromagnetic field is given by $\ket{\Psi_\text{cl}}=\mathcal{D}(\alpha_{\kappa})\ket{0}_\text{em}$, as defined in \eqnref{eq:Eclassical}, with the monochromatic coherent amplitude $\alpha_{\kappa}$ with frequency $\omega_0$. The output state is given by $\ket{\Psi^\text{S}_\text{out}}\equiv \hat{b}_\kappa^\dagger \hat{b}_\mu^\dagger \ket{\Psi_\text{in}}$ and  $\ket{\Psi^\text{aS}_\text{out}}\equiv \hat{b}_\kappa^\dagger \hat{b}_\mu \ket{\Psi_\text{in}}$ for the Stokes ($p=\text{S}$) and anti-Stokes ($p=\text{aS}$) transition, respectively. Here, $\hat{b}_\kappa^\dagger$ denotes the creation operator of a normalized plane-wave mode~\cite{Maurer2021}, which is the relevant mode for optical detection along a given direction $(\theta_k,\phi_k)$. The time-evolution operator in the Schr\"odinger picture is given by $\hat{U}(t,t')= \exp[- \im \Hop (t-t')/\hbar ]$, where $\Hop$ is our fundamental Hamiltonian~\eqnref{eq:QEDHamiltonianBosons}. We emphasize that for $\tau^p_{\kappa \mu}$ to be nonzero, the incoming photon in the Stokes or anti-Stokes transition, namely the photon in mode $\kappa'$ according to \figref{fig1}, has to be in a mode excited by the classical electromagnetic field, whereas the Stokes or anti-Stokes scattered photon in mode $\kappa$ can be in any of the available electromagnetic field modes, i.e. vacuum state or any mode excited by the electromagnetic field.

A quantity of interest in this context is the transition probability rate in the asymptotic limit~\cite{Cohen2004/2}  of the Stokes and anti-Stokes processes integrated over all $\kappa$, namely $\lim_{T\to \infty } \sum_\kappa|\tau_{\kappa\mu}^\text{p}|^2/T$. As we show in \appref{App:transition_amplitudes}, the transition probability rates can be evaluated using first order perturbation theory (i.e. Fermi's golden rule), and lead to
\begin{align}
  \lim_{T\to \infty } \frac{\sum_\kappa|\tau_{\kappa\mu}^\text{S}|^2}{T} & = (n_\mu+1) \Gamma^+_\mu, \\
    \lim_{T\to \infty } \frac{\sum_\kappa|\tau_{\kappa\mu}^\text{aS}|^2}{T} & = n_\mu \Gamma^-_\mu,
\end{align}
where
\be 
	\Gamma^{\pm}_\mu  \equiv 2\pi \sum_{\kappa} |G_{\kappa\mu}|^2 \delta(\omega_\kappa - \omega_0 \pm \Omega_\mu).
\ee 
The linearized couplings in free space are broadband over a frequency window given by $\Omega_\mu$, i.e. they fulfil  $|\Gamma_\mu^+ - \Gamma_\mu^-|  \ll \Gamma_\mu^+ + \Gamma_\mu^- $, and hence we define the transition probability rate
\be\label{eq:recoilheatinggeneral}
	\Gamma_\mu  = 2\pi \sum_{\kappa} |G_{\kappa\mu}|^2 \delta(\omega_\kappa - \omega_0).
\ee 
That is, $\Gamma_\mu \approx \Gamma^+_\mu \approx \Gamma^-_\mu$ in the broad coupling regime. In this regime one can show using standard techniques in open quantum system dynamics (e.g. Langevin equations, Born-Markov master equation) that the reduced motional dynamics described by  the fundamental Hamiltonian~\eqnref{eq:QEDHamiltonianBosons} lead to 
\be
\frac{d}{dt} \avg{ \bdop_\mu \bop_\mu} (t) = \Gamma_\mu,
\ee
that is, center-of-mass recoil heating with a phonon heating rate given by $\Gamma_\mu$~\cite{Jain2016}. Let us emphasize that when the particle is not in free-space but in environments that change the mode structure of the electromagnetic field  (e.g. a particle inside an optical resonator), the system is not in the broadband coupling regime, leading to  $\Gamma^+_\mu \neq \Gamma^-_\mu$, and hence to the possibility of passively cooling the center-of-mass motion of the particle~\cite{Ballestero2019,Windey2019,Delic2019,Delic2020,Ranfagni2022,Piotrowski2022}.

The photons scattered via a Stokes or anti-Stokes process carry information about the center-of-mass motion, that is, they are entangled with the center-of-mass motional degrees of freedom. The knowledge of where they scatter, namely their angular probability distribution, is key to collect them, measure them, and process the information to exert active feedback to the particle. This knowledge is provided by the information radiation pattern (IRP) $\mathcal{I}^p_\mu (\theta_k,\phi_k)$ associated to the center-of-mass motion along the $\mu$-axis, which is defined as the normalized angular distribution  of the transition probability rate in the asymptotic limit, namely
\be 
    \mathcal{I}^p_\mu(\theta_k,\phi_k)\equiv  \frac{\lim_{T\to \infty } \int_0^\infty dk k^2 \sum_g |\tau^p_{\kappa\mu}|^2}{\lim_{T\to \infty } \sum_\kappa |\tau^p_{\kappa\mu}|^2}.
\ee
Note that by definition $\int d\theta_k d\phi_k \sin\theta_k \mathcal{I}^p_\mu(\theta_k,\phi_k) = 1$. The IRP $\mathcal{I}^p_\mu (\theta_k,\phi_k)$ provides the solid-angle probability distribution of a photon scattered either through a Stokes or anti-Stokes process. Within the broadband coupling regime, one obtains that 
\begin{multline} \label{eq:informationRadiationPattern}
    \mathcal{I}_\mu(\theta_k,\phi_k)  =
    \\ \frac{  \sum_{g}\int_0^\infty dk k^2 |\sum_{\kappa'}S_{\kappa\kappa'}G_{\kappa'\mu}|^2 \delta(\omega_\kappa - \omega_0)}{\sum_\kappa |G_{\kappa\mu}|^2 \delta(\omega_\kappa - \omega_0)},
\end{multline}
where $S_{\kappa\kappa'}$ denotes the single-photon scattering matrix~\cite{Maurer2021}. In this regime, the IRP is the same for a Stokes and anti-Stokes process, showing that these processes can be distinguished by measuring the frequency of the scattered photon but not by its angular probability distribution.

Both the laser recoil heating rate $\Gamma_\mu$ and the IRP $\mathcal{I}_\mu(\theta_k,\phi_k)$ are fully determined by the linearized coupling rate $G_{\kappa\mu}$, that is,  by the coupling rates $g_{\kappa\kappa'\mu}$ (\eqnref{eq:gcouplings}), the coherent amplitude $\alpha_\kappa$ associated to a given classical electromagnetic field~(\eqnref{eq:Eclassical}), and the single-photon scattering matrix $S_{\kappa\kappa'}$. In \appref{App:explicit_IRP_recoil} we show how \eqnref{eq:recoilheatinggeneral} and \eqnref{eq:informationRadiationPattern} can be expressed in an explicit form as a sum over discrete angular momentum and polarization indices, see \eqnref{eq:recoil_explicit} and \eqnref{eq:IRPexplicit}, which makes their evaluation numerically efficient. Hence, our theory allows us to efficiently evaluate recoil heating rates and IRPs for spheres of arbitrary refractive index and size, and  in presence of arbitrary combinations of electromagnetic field modes in a coherent state. This is one of the main result of this article.

The knowledge of the laser recoil heating rate $\Gamma_\mu$ and the IRP $\mathcal{I}_\mu(\theta_k,\phi_k)$ is crucial to implement active feedback cooling via optical detection~\cite{Tebbenjohanns2019CD,Conangla2019,Magrini2021,Tebbenjohanns2021, Kamba2022}. In particular, the minimum achievable mean phonon-occupation number reads~\cite{Tebbenjohanns2019,Magrini2021,Tebbenjohanns2021}
\begin{equation}\label{eq:cond}
    \bar{n}_\mu \equiv \frac{1}{2}\left(\frac{1}{\sqrt{\eta_\mu}}-1\right).
\end{equation}
Here $\eta_\mu \equiv \eta_\mu^d \eta_\mu^e\eta_\mu^o$ denotes the total efficiency and is determined by: (i) The detection efficiency
\be \label{eq:etad}
\eta^d_\mu \equiv \int_{S_d} \text{d} \theta_k \text{d} \phi_k \sin \theta_k \mathcal{I}_\mu(\theta_k,\phi_k),
\ee 
where $S_d$ denotes the solid angle that is covered by the collection lens. (ii) The efficiency associated to environmental information loss 
\begin{equation} \label{eq:barn}
    \eta^e_\mu \equiv \frac{\Gamma_\mu}{\Gamma_\mu + \Gamma_{\mu}^g},
\end{equation}
where the heating rate $\Gamma_\mu^g$ due to environmental gas at pressure $p$ and temperature $T$ is given by
\begin{equation} \label{eq:Gammagas}
    \Gamma_\mu^g=0.619 \frac{r_{0\mu}^2 p  }{\hbar^2} 6\pi R^2 \sqrt{\frac{8 k_B T m_0}{\pi }}.
\end{equation}
Here $m_0$ denotes the molecular mass of the environmental gas and $k_B$ is the Boltzmann constant~\cite{Beresnev1990,Li2011}. Note that we are assuming that other sources of environmental noise (e.g. displacement noise in the trapping potential, center-of-mass coupling to internal acoustic phonons, and emission of black-body radiation) are negligible, namely their associated heating rates are smaller than $\Gamma_\mu + \Gamma_\text{g}$. (iii) The efficiency $\eta_\mu^o$ associated to all other information loss channels in the active feedback scheme (e.g. mode-matching, detector noise, digital noise)~\cite{Magrini2021,Tebbenjohanns2021}. Hereafter, we consider ideal feedback, i.e. $\eta_\mu^o = 1$. 

\eqnref{eq:cond} shows that a necessary condition to achieve center-of-mass ground-state cooling along the $\mu$-axis, defined as  $\bar n_\mu <1$, is $\eta_\mu >1/9 \approx 0.11$. We emphasize that our analysis will be particularly relevant in experimental situations where center-of-mass heating rates are dominated by $\Gamma_\mu + \Gamma_\text{g}$ and for nearly ideal feedback schemes with $\eta_\mu^o \lesssim 1$. Finally, we remark that with particles beyond the point-dipole approximation, recoil heating rates will be typically comparable or even larger than the mechanical frequencies, which will require fast feedback schemes optimized to operate in these regimes.

In the following section, we will explicitly show how this theory can be applied to study a case which is very relevant to current experimental efforts~\cite{Tebbenjohanns2019CD,Conangla2019,Magrini2021,Ciampini2021,Tebbenjohanns2021,Kamba2021,Kamba2022}: a dielectric silica sphere of arbitrary refractive index and size interacting with a focused monochromatic laser, either in a running or standing-wave configuration. Our analysis will show in which regimes center-of-mass ground-state cooling of a sphere could be achieved via active feedback cooling, that is, in which regimes $ \eta_\mu^d \eta_\mu^e >1/9$ and hence $\bar n_\mu < 1$. Remarkably, we will predict several experimentally feasible configurations in which simultaneous 3D ground-state cooling via active feedback is possible.

\begin{figure}[t]
    \centering
    \includegraphics[width=0.9\columnwidth]{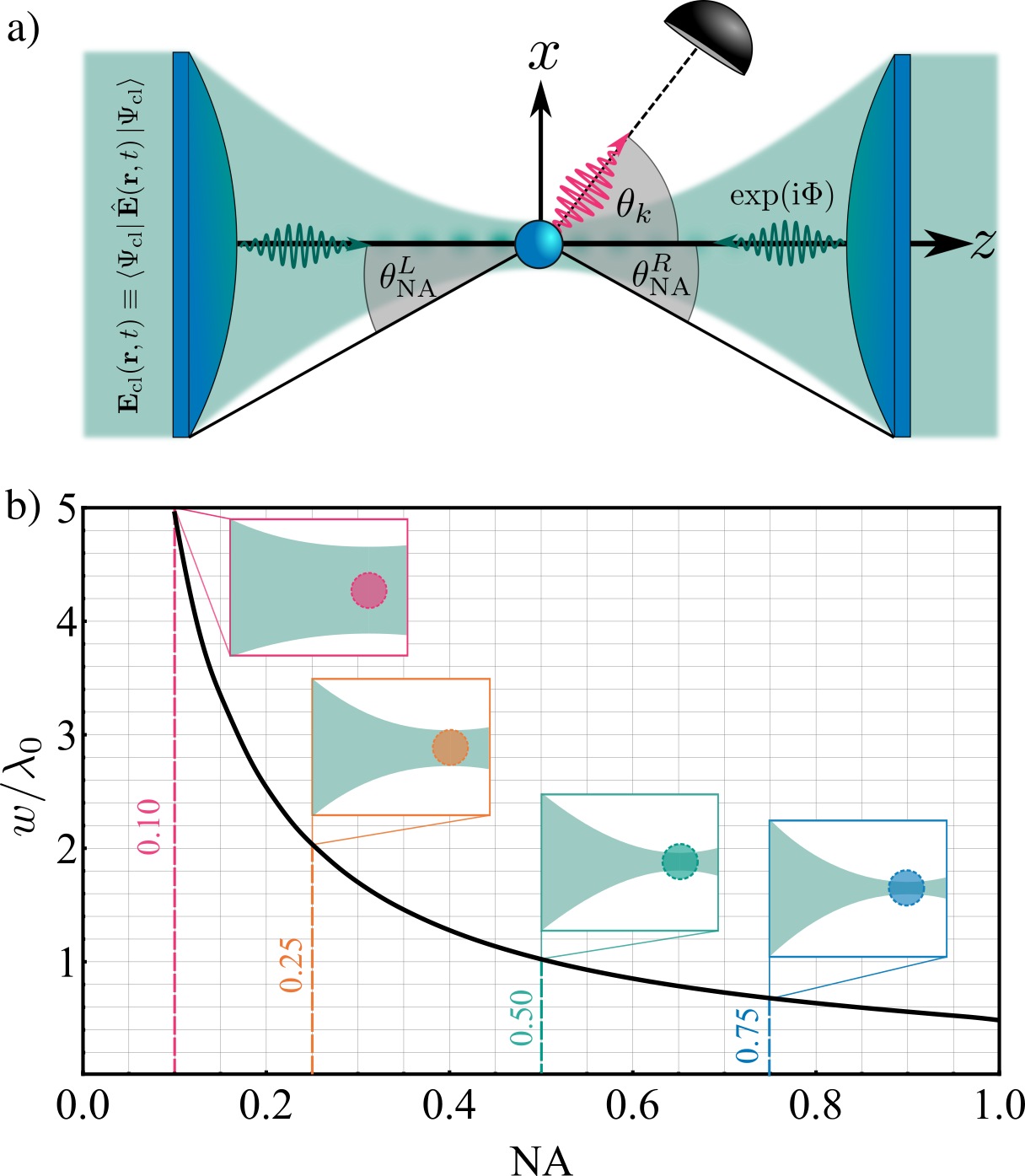}
    \caption{a) Sketch of the optical configuration. We consider one or two incoming $x$-polarized monochromatic Gaussian beams of frequency $\omega_0$ propagating along the positive and negative $z-$axis. The beams are focused by an aplanatic lens which are assumed to be overfilled in order to achieve strong focusing at the focal point, where a dielectric sphere of radius $R$ is placed. In the two-beam setup $\Phi$ denotes the relative phase between the two counter-propagating beams. The IRPs $\mathcal{I}_\mu(\theta_k,\phi_k)$ are defined with respect to the same coordinate system, where $\theta_k$ denotes the angle between the positive $z-$axis and the detector. b) Waist $w$, as a function of the numerical aperture NA of the lens. The insets show a sketch of the relative size of the sphere and the waist of the beam for $R/\lambda_0 = 1$.}
    \label{fig2}
\end{figure}

\section{Case study} \label{sec:case_study}

Let us apply the theoretical methods developed in this article to study a case that is of relevance to current experimental efforts. In particular, we will evaluate the recoil heating rates $\Gamma_\mu$ and the information radiation patterns $\mathcal{I}_\mu(\theta_k,\phi_k)$ for a trapped dielectric sphere made of silica interacting with a focused laser beam, either in a running wave configuration or in a standing-wave configuration, see \figref{fig2}. In both cases the equilibrium position of the dielectric sphere is assumed to be at the focus. In the standing-wave configuration, we will include a relative phase $\Phi$ between the two counter-propagating beams, so that at focus one can have an intensity maximum ($\Phi=0$), an intensity gradient maximum ($\Phi = \pi/2$), or an intensity minimum ($\Phi = \pi$). In addition, we will evaluate the detection efficiency $\eta_\mu^d$ \eqnref{eq:etad} and the phonon mean number occupation $\bar n_\mu$ \eqnref{eq:barn} for backward and forward detection and ideal feedback. These quantities will unveil in which situations ground-state cooling of 1D, 2D, and even 3D center-of-mass motion is possible. 

We remind the reader (see \eqnref{eq:potential_condition}) that in this article we assume that 3D harmonic trapping with an equilibrium position at the focus with trap frequencies given by $\Omega_\mu$ is possible when taking into account the optical forces generated by the laser beams. We do not specify, however, how this is implemented, i.e. whether using all-optical and gravitational forces~\cite{Ashkin1977,Li2011,BlakemorePRA2019,MonteiroPRA2020,Arita2022}, hybrid schemes combining low frequency electric and optical forces~\cite{Millen2015,Conangla2020,Lorenzo2021}, etc. In this way our analysis is very general, it even includes non-levitating scenarios, e.g. a dielectric particle attached to a cantilever. We also remark that the IRPs and recoil heating rates calculated in this section are also relevant in situations where the particle is not harmonically trapped (e.g. in an inverted harmonic potential~\cite{Romero-Isart2017,Pino2018,Weiss2021,Neumeier2022}). This is the case provided the sphere's center-of-mass fluctuations are small enough to justify the coupling linear with the center-of-mass position used at the starting point of the theory, see \eqnref{eq:QEDHamiltonian}.

\subsection{Coherent amplitudes}

The first task to apply our theory is to derive an expression for the coherent amplitudes $\alpha_\kappa$ associated to the optical configuration we are considering, see \figref{fig2}.  This is done by first deriving an expression for the classical electric field and then inferring $\alpha_\kappa$ through \eqnref{eq:classicalField}.  For notational convenience, we will label the lens placed at negative $z$ values as left (L), and the lens placed at positive $z$ values as right (R). The lenses can be used to focus and/or collect the scattered light. 
We derive the classical electric field  of an incoming monochromatic $x-$polarized Gaussian beam $\mathbf{E}_\text{in}(\rr,t) = \text{Re}[\mathbf{E}_\text{in}(\rr)\exp(-\im \omega_0 t)]$ with mode profile $\mathbf{E}_\text{in}(\rr)$ and frequency $\omega_0$ focused by an aplanatic lens, see \figref{fig2} and \appref{App:gaussian} for details. First, we focus on a single Gaussian beam propagating along the positive $z-$axis (from the left) and whose focus coincides with the origin of coordinates (i.e. with the equilibrium position of the dielectric sphere). As shown in \appref{App:gaussian}, the corresponding coherent amplitudes $\alpha_\kappa^L$ for each polarization $g=1,2$ read
\begin{multline}
    \left[
    \begin{array}{c}
         \alpha_1^L(k,\theta_k,\phi_k)  \\
         \alpha_2^L(k,\theta_k,\phi_k) 
    \end{array}
    \right] = \sqrt{\frac{4\pi k_0 P \vert \cos\theta_k\vert}{\hbar c^2 \sin^2\theta_{\rm NA}^L}}\frac{\delta(k-k_0)}{k_0^2}
    \\
    \times \exp\left(-\frac{\sin^2\theta_k}{f_0^2\sin^2\theta_{\rm NA}^L}\right)
    \left[
    \begin{array}{c}
         \im\sin\phi_k  \\
         \cos\phi_k 
    \end{array}
    \right],
\end{multline}
where $k_0 = 2\pi/\lambda_0 \equiv \omega_0/c $. The parameters that determine the coherent amplitude are the optical power $P$ of the focused field, the numerical aperture $\text{NA}=\sin\theta_\text{NA}$ of the lens, and the filling factor $f_0$. We assume that the field overfills the lens, $f_0 \gg 1$, to guarantee maximum focusing of the beam~\cite{novotny_hecht_2012}.  A similar calculation enables to compute the amplitudes for a beam propagating along the negative $z-$axis (from the right), which can be written as $\alpha_1^R(k,\theta_k,\phi_k) = \alpha_1^L(k,\theta_k,\phi_k)$ and $\alpha_2^R(k,\theta_k,\phi_k) = -\alpha_2^L(k,\theta_k,\phi_k)$. By combining the amplitudes of the left and right incoming beams and adding a relative phase $\Phi$, we obtain the coherent amplitude corresponding to two counter-propagating focused Gaussian beams in a standing wave configuration, i.e. $\alpha_\kappa^L + \alpha_\kappa^R \exp(\im\Phi)$. Here, the phase $\Phi$ determines the intensity that scales as $\cos^2(\Phi/2)$  at the particle position, so that it lies at a node and anti-node for $\Phi=\pi$ and $\Phi = 0$, respectively.  In \figref{fig2}b, we show how the waist $w$ of the focused Gaussian beam depends on the numerical aperture of the overfilled lens, see \appref{app:charactFocusField} for details. As expected the waist decreases rapidly with increasing numerical aperture and reaches a value on the order of the wavelength $\lambda_0$ which corresponds to the diffraction limit~\cite{novotny_hecht_2012}.

With the determined expression of the coherent amplitudes $\alpha_\kappa$ for the optical configuration of \figref{fig2}, we can now calculate the recoil heating rates $\Gamma_\mu$~\eqnref{eq:recoilheatinggeneral}, the IRPs $\mathcal{I}_\mu(\theta_k,\phi_k)$~\eqnref{eq:informationRadiationPattern}, the detection efficiencies $\eta_\mu^d$~\eqnref{eq:etad} using the left or the right lens to collect the light, and the corresponding phonon mean-number occupation $\bar n_\mu$~\eqnref{eq:barn}. To do so, we will consider the physical parameters listed in Tab. \ref{tab:parameters}. 

\begin{table}
	\begin{tabularx}{0.9\columnwidth}{l r}
		\toprule
		\textbf{Parameters} & \text{Description }\\
		\midrule
		\midrule
		$\epsilon = 2.07$ & relative permittivity (silica)~\cite{Malitson1965} \\
        $\rho = 2200 \text{ kg}/\text{m}^3$ & mass density (silica)~\cite{Bass2001} \\
		$\lambda_0 = 1550\text{ nm}$ & laser wavelength \\
		$P =200 \text{ mW}$ & power per laser beam \\
		$f_0 = 10$ & lens filling factor \\
        $m_0 = 4.81\times 10^{-26}\text{kg}$ & molecular mass (air) \\
        $p = 10^{-9}$ \text{mbar} & gas pressure \\
        $T = 300 \text{K}$ & gas temperature \\
		\bottomrule
	\end{tabularx}
	\caption{Parameters considered throughout this article.}
	\label{tab:parameters}
\end{table}

\subsection{Single focused beam}
\label{sec:runningWave}

\begin{figure}[t!]
    \centering
    \includegraphics[width=\columnwidth]{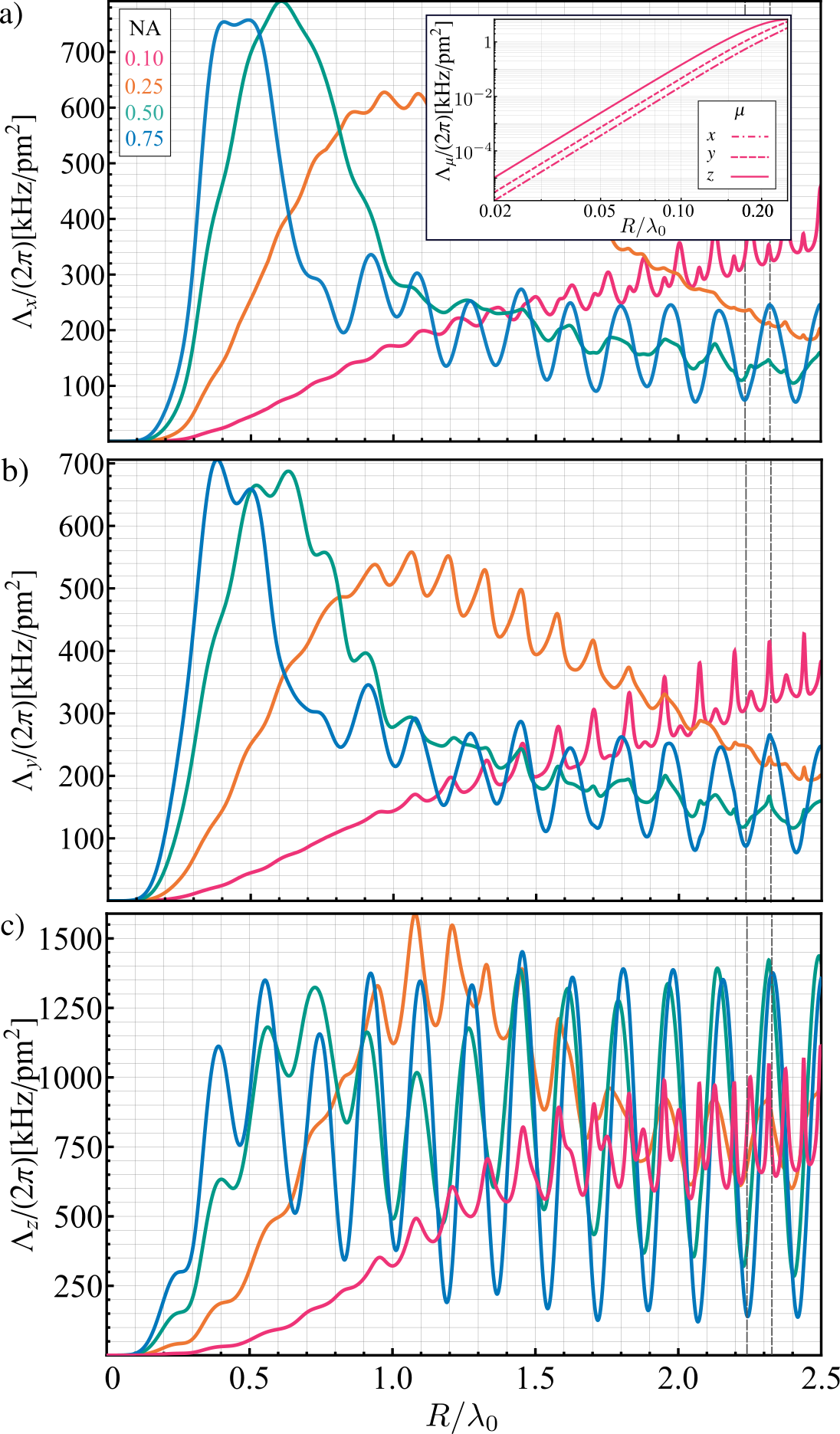}
    \caption{\LambdaNameCapitalized $\Lambda_\mu = \Gamma_\mu / r_{0\mu}^2$ for a single focused $x-$polarized Gaussian beam propagating along the positive $z-$axis as a function of the silica sphere's radius $R/\lambda_0$ and the numerical aperture of the lens NA $= 0.10$, $0.25$, $0.50$, $0.75$, and for all three axes $\mu=x,y,z$ in panel a-c respectively. The values for the power, wavelength, and relative permittivity are listed in Tab.~\ref{tab:parameters}. The inset in panel a shows a detailed view of the small-particle regime for all three axes at $\text{NA}=0.10$ in a log-log plot. The vertical grey dashed lines specify the radii for which the IRPs are shown in the last two columns in \figref{fig4}.}
    \label{fig3}
\end{figure}

Let us first analyse the results for the running wave configuration, where the left lens in \figref{fig2} acts as a focusing and collection lens while the right lens acts as a collection lens only. In \figref{fig3} we show $\Lambda_\mu \equiv \Gamma_\mu / r_{0\mu}^2 $, which we hereafter call recoil localizaton parameter~\footnote{In the context of position localization decoherence, the localization parameter is very relevant to describe the decoherence rate of quantum states that are spatially delocalized over scales larger than the zero-point motion~\cite{Joos1985,Schlosshauer2007,Romero-Isart2011QS}.}. The \LambdaName is the recoil heating rate divided by the squared zero-point motion and it hence does not depend on the trapping frequency $\Omega_\mu$, c.f. \eqnref{eq:recoil_explicit} in \appref{App:explicit_IRP_recoil}. Moreover, since $\Gamma_\mu^g$ is also proportional to $r_{0 \mu}^2$, see \eqnref{eq:Gammagas}, the phonon mean number occupation $\bar n_\mu$~\eqnref{eq:barn} does also not depend on $\Omega_\mu$. 

\figref{fig3} shows the \LambdaName for various numerical apertures NA and radii $R$ ranging from the small-particle regime $R/\lambda_0 \ll 1$ into the Lorenz-Mie regime $R/\lambda_0 > 1$. The inset in \figref{fig3}a shows the small-particle regime for all three axes and $\text{NA}=0.1$ in greater detail. As expected, in the small-particle regime we recover the results obtained using the coupling rates in the point-dipole approximation (\eqnref{eq:gcouplings_small_particle}), where the \LambdaName scales as $\Lambda_\mu\propto (R/\lambda_0)^6$~\cite{Ballestero2019, Seberson2020}. Beyond the small-particle regime, $\Lambda_\mu$ deviates from this  polynomial scaling in size and transitions into an oscillatory behaviour after reaching a  maximum. The first maximum, which is not always global, is reached at smaller radii as one considers larger numerical apertures. The oscillatory features are most pronounced for $\mu = z$ and $\text{NA}=0.75$ and can be understood intuitively using the following simple model (c.f. \appref{App:slab}). For large numerical apertures the waist of the focused beam is on the order of the wavelength (see \figref{fig2}b). For spheres of radii $R/\lambda_0 \gtrsim 1$ the system can therefore be modelled by a collimated beam that is normally incident on a dielectric slab of thickness $2R$, i.e. a Fabry-Pérot interferometer. Note that this regime is very different from the small-particle regime or from the case of small numerical apertures, where a non-vanishing intensity is incident across the whole surface of the sphere. Using this one-dimensional toy model, involving only two electromagnetic modes, one can show that the \LambdaName can be expressed in terms of the reflectance and transmittance of the interferometer, see \appref{App:slab}. These quantities oscillate with a period $(4\sqrt{\epsilon})^{-1}$ in $R/\lambda_0$, which agrees with the oscillations observed in \figref{fig3}. Note that the typical values for zero-point motion of particles beyond the small-particle limit are in the picometer scale, and hence \figref{fig3} shows that recoil heating rates $\Gamma_\mu$ of several hundreds of  kilohertz are expected. These recoil heating rates will be typically comparable and larger than typical mechanical frequencies, and hence feedback cooling schemes should be optimized to operate in these regimes.

\begin{figure*}
    \centering
    \includegraphics[width=0.8\textwidth]{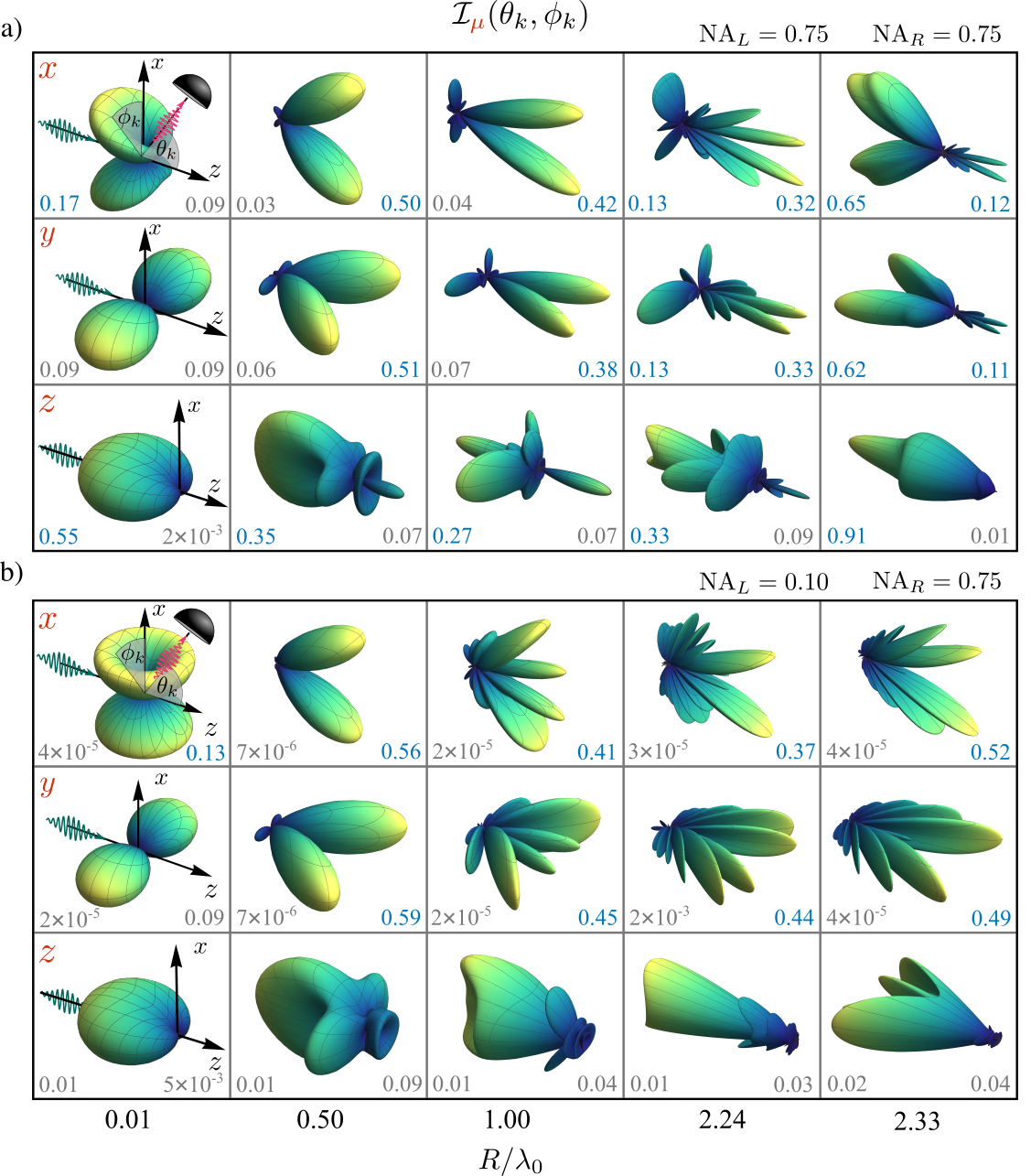}
    \caption{Information radiation patterns $\mathcal{I}_\mu(\theta_k,\phi_k)$ of a silica sphere and a single focused $x-$polarized Gaussian beam propagating along the positive $z-$axis (reference frame in first row). The value of the IRP is encoded both in the radial distance from the center and the color scale. The focusing lens has a numerical aperture $\text{NA}_L =0.75$ in panel a and $\text{NA}_L =0.10$ in panel b, while the collection lens has a numerical aperture $\text{NA}_R =0.75$ in both panels. The detection efficiency for the focusing lens and collection lens is shown in each sub-panel (highlighted in blue for $\eta_\mu^d > 1/9$). Across panels the value of $R/\lambda_0$ is, for each column, constant and indicated below the last row.}
    \label{fig4}
\end{figure*}

In \figref{fig4} we show the IRP for all three axes, a left lens with numerical aperture $\text{NA}_L=0.75$ in \figref{fig4}a and $\text{NA}_L=0.10$ in \figref{fig4}b, a right lens with numerical aperture $\text{NA}_R=0.75$ in both panels, and a wide range of particle radii. Each sub-panel contains the detection efficiency $\eta_\mu^d$ (\eqnref{eq:etad}) for the left lens (lower left corner) and right lens (lower right corner). All values for which the necessary condition for ground-state cooling, i.e. $\eta_\mu^d > 1/9$ is met, are highlighted in blue (dark grey). The first column corresponds to the known IRPs~\cite{Tebbenjohanns2019} in the small-particle regime where almost all photons carrying information about the center-of-mass displacement along $z$ are backscattered leading to a large detection efficiency $\mu_z^d= 0.55 \gg 1/9$ for the high-NA configuration. The knowledge of the IRP in the small-particle regime has recently enabled the achievement of ground-state cooling along the $z$-axis via active feedback~\cite{Magrini2021,Tebbenjohanns2021}.
While for $R/\lambda_0 \leq 0.5$ the IRPs in the low- and high-NA configuration are very similar, we observe that, for larger radii, their shape strongly depends on the numerical aperture $\text{NA}_L$ of the focusing lens (left lens). For $\text{NA}_L=0.1$ the IRPs along $\mu =x,y$ become increasingly sharply peaked in the forward direction as the radius increases. In contrast to the small-particle regime, the detection efficiency in the forward direction is orders of magnitude larger than the backward direction and exceeds the ground-state threshold of $1/9$. For $\text{NA}_L=0.75$ the angular distribution of the IRPs show more complex features. Note that for $R/\lambda_0 = 2.24 \text{ and } 2.33$ we observe a local minimum and maximum for $\Lambda_\mu$ and NA$=0.75$ along all three axes (recall gray dashed lines in \figref{fig3}). In the corresponding IRPs in \figref{fig4}a we can see how a large \LambdaName goes hand in hand with a large amount of back-scattered photons, i.e. large reflectance in the Fabry-Pérot interferometer. For the $z-$axes this leads to very high detection efficiencies of up to $\eta_z^d\gtrsim0.9$. As we show below, this trend is general since we obtain detection efficiencies that exceed the ground-state threshold for many radii in all other axes and for both forward- and backward detection.

\begin{figure}[t!]
    \centering
    \includegraphics[width=0.95\columnwidth]{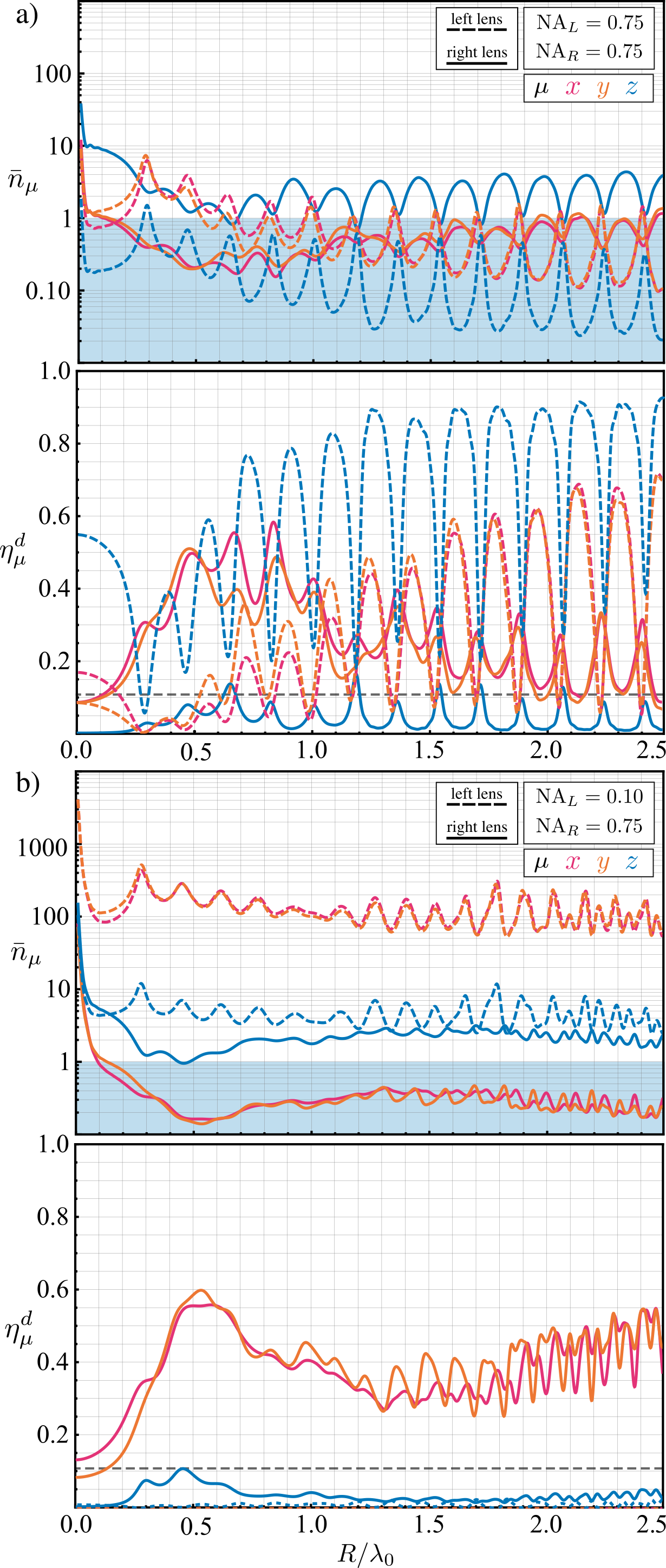}
    \caption{Detection efficiencies $\eta_\mu^d$ and minimum achievable mean phonon occupation number $\bar{n}_\mu$ along $x$, $y$, and $z$ for ideal feedback as a function of $R/\lambda_0$ and $p = 10^{-9}\text{mbar}$, $T=300\text{K}$, at the same optical configuration as in \figref{fig3}. The dashed (solid) lines correspond to the values at the left (right) lens. The blue shaded area highlights the region where $\bar{n}_\mu < 1$ and the grey dashed line shows $\eta_\mu^d = 1/9$ respectively.}
    \label{fig5}
\end{figure}

In \figref{fig5} we show the detection efficiencies $\eta_\mu^d$ at the left lens (dashed lines) and right lens (solid lines) as a function of the radius for all three axes. The threshold  $\eta_\mu^d= 1/9$ is indicated with a horizontal dashed grey line. These detection efficiencies are then combined with the efficiency $\eta_\mu^e$ associated to environmental information loss due to gas scattering (\eqnref{eq:etad}) to obtain the minimum achievable mean phonon occupation number $\bar{n}_\mu$ for ideal feedback (\eqnref{eq:barn}) for the physical parameters given in Tab.~\ref{tab:parameters}. The blue-shaded area highlights the regions for which $\bar{n}_\mu < 1$. \figref{fig5}a shows how a high detection efficiency $\eta_z^d$ in the backward direction (blue, dashed, highest curve at $R/\lambda_0 \to 0$) enabled the experimental realization of center-of-mass ground-state cooling using shot-noise limited optical detection and active feedback~\cite{Magrini2021,Tebbenjohanns2021}. While for $\mu = x,y$ the detection efficiency in the small-particle limit is much smaller, they reach comparably large values in the forward direction (solid) at $R/\lambda_0 \simeq 0.5$. The same can be said for the backward direction (dashed) for $R/\lambda_0 > 1.0$, where one again encounters the periodic features of the recoil localization parameter. Remarkably, this shows that there exists a broad range of parameters where simultaneous three-dimensional ground-state cooling is possible for the high-NA configuration. Similar conclusions can be drawn for the the low-NA setup ($\text{NA}_L=0.10$ and  $\text{NA}_R=0.75$) in \figref{fig5}. We observe how the detection efficiency in the forward direction (solid) quickly grows with increasing radius. In fact the values oscillate around a value so large that three-dimensional ground-state cooling is also within reach for all three axes and $R/\lambda_0 > 0.5$.

\subsection{Standing wave}
\label{sec:standingWave}

\begin{figure}[t!]
    \centering
    \includegraphics[width=\columnwidth]{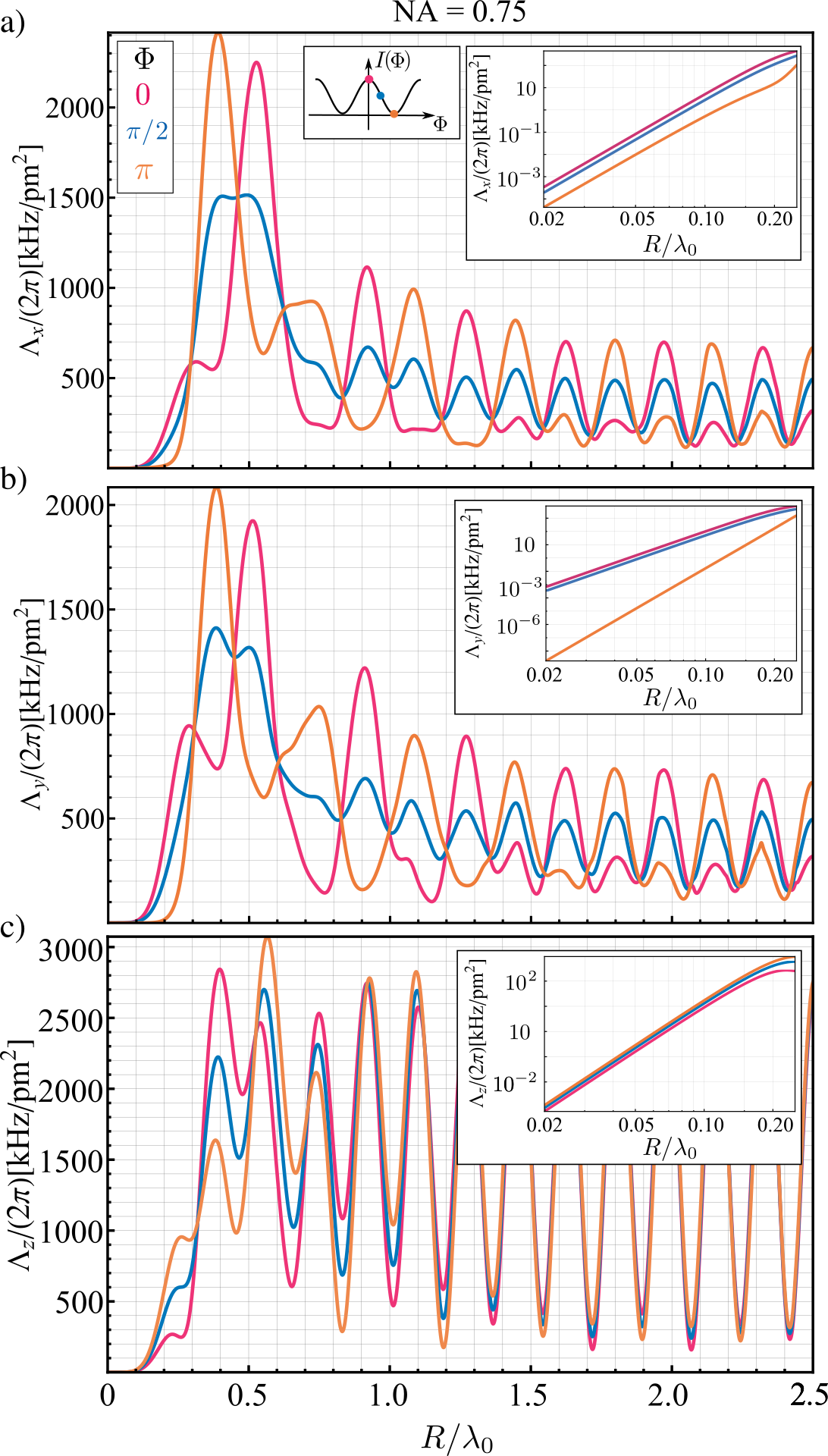}
    \caption{\LambdaNameCapitalized $\Lambda_\mu = \Gamma_\mu / r_{0\mu}^2$ for two focused $x-$polarized Gaussian beam counter-propagating along the $z-$axis as a function of the silica sphere's radius $R/\lambda_0$ and for all three axes $\mu=x,y,z$ in panel a)-c) respectively. The values for the power, wavelength, and relative permittivity are listed in Tab.~\ref{tab:parameters}.  Each panel shows the \LambdaName for $\Phi= 0$, $\pi/2$, $\pi$, $\text{NA}=0.75$, and an inset with a detailed view of the small-particle regime for all three relative phases in a log-log plot. Inset in panel a) maps the relative phase $\Phi$ to the corresponding intensity at the origin. }
    \label{fig6}
\end{figure}

Let us now analyze the results for two focused $x-$polarized Gaussian beams counter-propagating along the $z-$axis, where the lenses in \figref{fig2} act both as focusing and collection lenses. Here we focus on the high-NA case, namely $\text{NA}_L = \text{NA}_R =0.75$. The results for $\text{NA}_L = \text{NA}_R =0.10$ are shown in \appref{app:figures}.  

In \figref{fig6} we show the \LambdaName for three values of the relative phase $\Phi = 0,\pi/2,\pi$ (particle at the intensity maximum, intensity gradient maximum, and  intensity minimum, respectively) as as function of $R/\lambda_0$ ranging from the small-particle regime to the Lorenz-Mie regime for all three axes. Each panel contains an inset that shows the small-particle regime in a log-log plot in greater detail. \figref{fig6} shares many features with \figref{fig3}. The recoil localization parameters reach a maximum before they transition into an oscillatory behaviour that is most pronounced for $\mu = z$, where the lines for the different relative phases overlap (c.f.~\appref{App:slab} for an explanation in terms of the model based on the Fabry-Pérot interferometer). Let us remark that within the point-dipole approximation and low-NA regime (e.g. two counterpropagating plane waves), the recoil heating rates have some features that have been discussed in the literature: (i) the sum $\sum_{\mu} \Lambda_\mu$ is independent on $\Phi$ (namely the position of the particle in the standing wave)~\cite{Gordon1980} (ii) and $\Lambda_x= \Lambda_y=0$ at $\Phi=\pi$ (intensity maximum)~\cite{Neumeier2022}. Our theoretical treatment goes beyond the point-dipole approximation and low-NA assumptions, and hence shows features beyond (i) and (ii). For instance, for low-NA (see \figref{fig9} in \appref{app:figures}) we observe the expected $\Lambda_z \propto (R/\lambda_0)^6$ scaling that can be derived using the coupling rates in the point-dipole approximation. However, for $\mu=x,y$ and $\Phi = \pi$ we observe a polynomial scaling of higher order, namely $\Lambda_{x,y}\propto (R/\lambda_0)^{10}$. This is consistent with the fact that the coupling rates in the point-dipole approximation vanish when evaluated at this point, and as a consequence the lowest-order term is the next non-vanishing term in the small-particle expansion.  

\begin{figure*}
    \centering
    \includegraphics[width=0.7\textwidth]{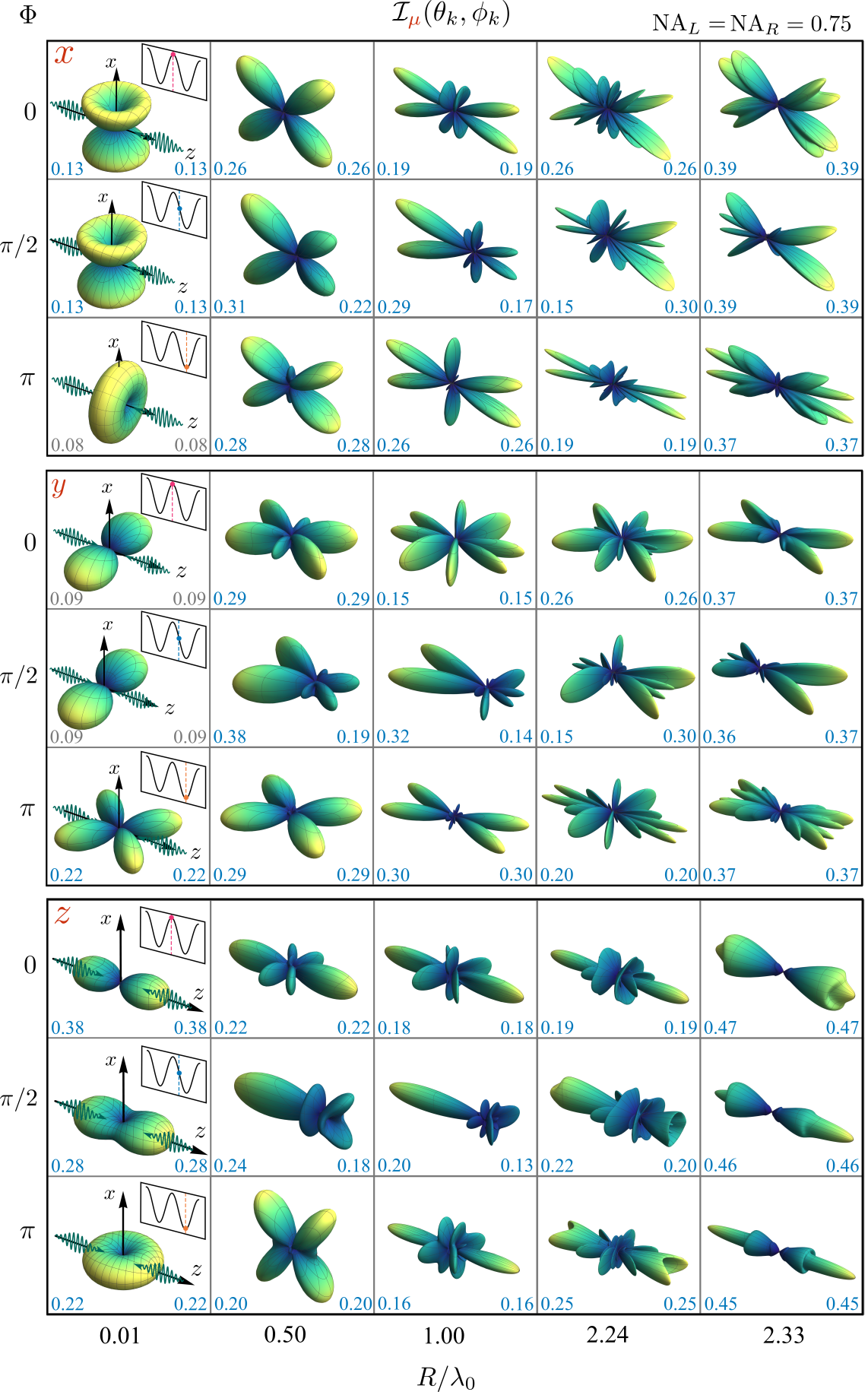}
    \caption{Information radiation patterns $\mathcal{I}_z(\theta_k,\phi_k)$ of a silica sphere, two focused $x-$polarized Gaussian beams counter-propagating parallel to the $z-$axis (reference frame in first row), and relative phases $\Phi=0,\pi/2,\pi$ (corresponding intensity at origin shown as inset in first column). The value of the IRP is encoded both in the radial distance from the center and the color scale. The two focusing lenses have a numerical aperture $\text{NA}_L=\text{NA}_R = 0.75$.  The detection efficiency for the left and right lens shown in each sub-panel (highlighted in blue for $\eta_\mu^d > 1/9$). Across panels the value of $R/\lambda_0$ is, for each column, constant and indicated below the last row.}
    \label{fig7}
\end{figure*}

\begin{figure*}
    \centering
    \includegraphics[width=0.8\textwidth]{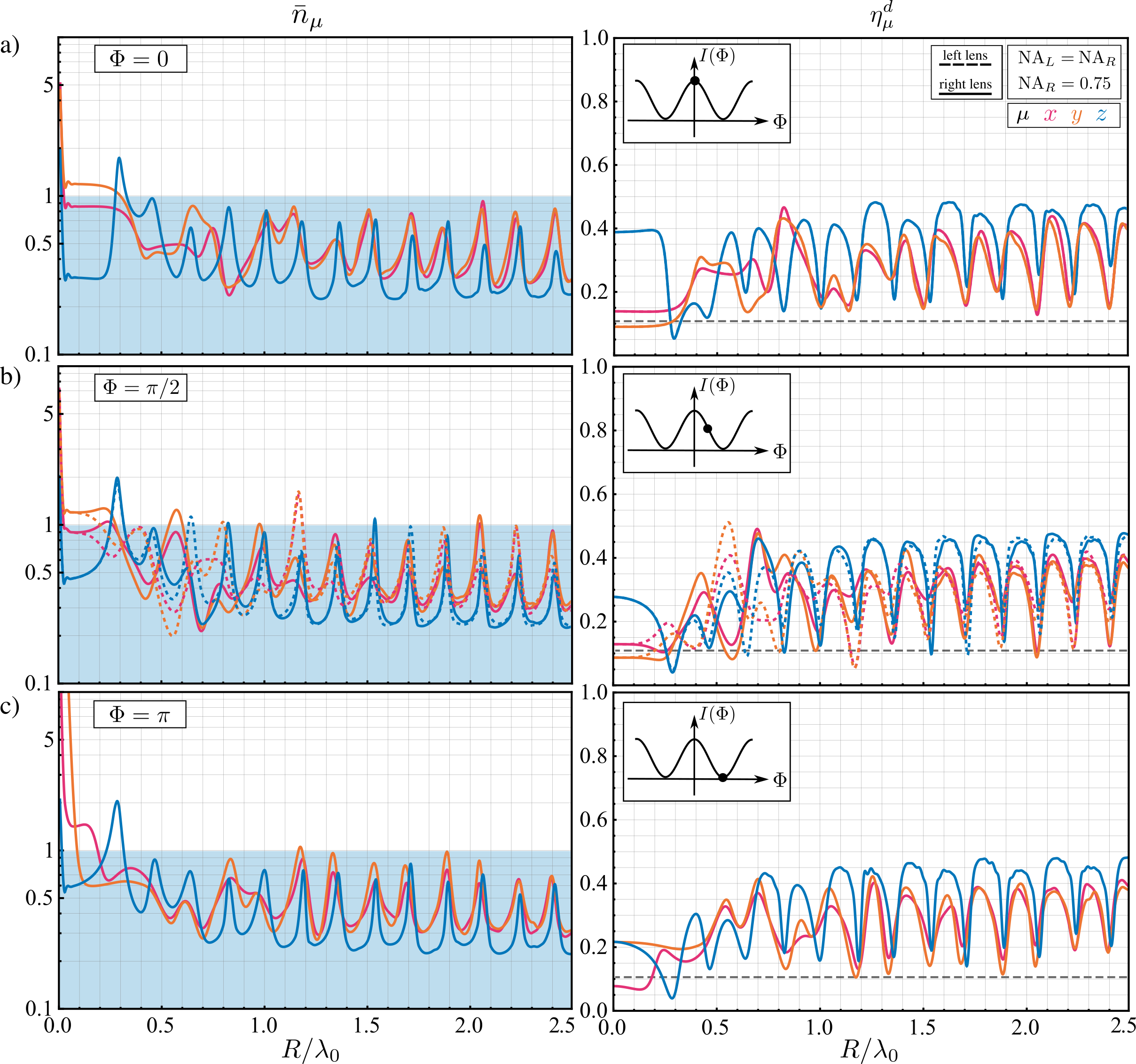}
    \caption{Detection efficiencies $\eta_\mu^d$ and minimum achievable mean phonon occupation number $\bar{n}_\mu$ along $x$, $y$, $z$ for ideal feedback as a function of $R/\lambda_0$ and $p = 10^{-9}\text{mbar}$, $T=300\text{K}$, at the same optical configuration as in \figref{fig7}, where panels a)-c) correspond to $\Phi = 0, \pi/2, \pi$ with an inset that maps the relative phase $\Phi$ to the corresponding intensity at the origin. The dashed (solid) lines correspond to the values at the left (right) lens. The blue shaded area highlights the region where $\bar{n}_\mu < 1$ and the grey dashed line shows $\eta_\mu^d = 1/9$ respectively.}
    \label{fig8}
\end{figure*}

In \figref{fig7} we show the IRPs. For $\Phi = 0, \pi$ the IRP are bound to be symmetric due to the symmetry of the problem. This is not the case for $\Phi = \pi/2$, where the IRPs can show large asymmetries as shown by the detection efficiency values in the lower left and right corner. Similar to the running-wave configuration for $\text{NA}_L = 0.75$ in \figref{fig4}a, the IRPs show a complex angular distribution and the ground-state threshold detection efficiency $\eta_\mu^d>1/9$ is reached for almost all axes, relative phases and radii under consideration. This is explicitly shown in \figref{fig8}, where we show the detection efficiencies at the left (dashed lines) and right (solid lines) lens as a function of the radius for all axes and relative phases ($\Phi = 0, \pi/2 , \pi$ in  a, b, c). The detection efficiencies are then combined with the efficiency $\eta_\mu^e$ associated to environmental information loss due to gas scattering (\eqnref{eq:etad}) to obtain the minimum achievable mean phonon occupation number $\bar{n}_\mu$ for ideal feedback (\eqnref{eq:barn}), and the gas pressure and temperature in Tab. \ref{tab:parameters}. While in the small-particle regime the detection efficiency along $x$ and $y$ lies below threshold for $\Phi = \pi$ and $\Phi = 0, \pi/2$ respectively, we see that for almost all the remaining parameters 3D ground-state cooling is possible for a broad range of parameters. Finally let us emphasize that the IRPs and recoil heating rates (also called back-action noise rates) obtained at $\Phi=\pi$ (intensity minimum) are particularly relevant in scenarios where the inverted harmonic optical potential is used for exponentially expanding the center-of-mass position probability distribution, which is of interest for enhancing optical position detection~\cite{Romero-Isart2017,Pino2018,Neumeier2022}.

\section{Conclusions} \label{Sec:Conclusions}

In the first part of this article we have developed a quantum theory of light interacting with the center-of-mass degrees of freedom of a dielectric sphere of arbitrary refractive index and size. The theory assumes the fluctuations of the center-of-mass to be small enough so that the light-matter coupling is linear in the center-of-mass position. This theory makes use of the quantization of the electromagnetic field in the presence of a  dielectric sphere of arbitrary refractive index and size, a task thoroughly derived in~\cite{Maurer2021} based on~\cite{Glauber1991}. This point is key, as the use of normalized scattering eigenmodes, as opposed to plane-wave modes, clearly unveil the Stokes and anti-Stokes processes which are responsible for describing the optomechanical physics of the problem. Furthermore, the spherical shape of the dielectric object allows us to perform an analytical treatment using many of the available analytical tools in spherical coordinates, thereby leading to optomechanical coupling rates, recoil heating rates, and information radiation patterns, that can be efficiently evaluated. We emphasize however, that we expect the theoretical methods of this article to be extendable to other shapes and other degrees of freedom (e.g. rotations~\cite{Arita2013,Kuhn2015,Hoang2016,Kuhn2017,Kuhn2017/2,Monteiro2018,Rashid2018,Ahn2018,Ahn2020,Laan2020,Pontin2022}, internal acoustic phonons~\cite{Hummer2020, Ballestero2020,Ballestero2020/2,Henkel2022}). We expect the recipe to be the same: (i)~Quantize the electromagnetic field with the object at equilibrium (c.f.~\cite{Maurer2021}). (ii)~Obtain the equations of motion of the relevant degrees of freedom driven by an electromagnetic force / torque (c.f.~\eqnref{eq:eom_RP}) which can be derived from a Hamiltonian with a linear coupling to the relevant degrees of freedom (c.f.~\eqnref{eq:QEDHamiltonian}). (iii)~Using classical electrodynamics express the electromagnetic force / torque acting on the degrees of freedom of interest written in terms of the electromagnetic fields (c.f.~\eqnref{eq:definition_F}). (iv)~Introduce the quantized electromagnetic fields in the presence of the object at equilibrium (c.f.~\eqnref{eq:electric_operator_eigenmodes} and \eqnref{eq:magnetic_operator_eigenmodes}) into the initial Hamiltonian to obtain a Hamiltonian describing the interaction between bosonics modes (c.f.~\eqnref{eq:QEDHamiltonianBosons}) and the corresponding optomechanical couplings (c.f.~\eqnref{eq:gcouplings}), and (v)~Linearize the theory using a classical electromagnetic field of relevance (c.f.~\eqnref{eq:classicalField}) to obtain a quadratic Hamiltonian (c.f.~\eqnref{eq:linearizedHamiltonian}) that can be used to evaluate recoil heating rates and information radiation patterns.

In the second part of this article, we have shown how the developed theory can be used to evaluate the recoil heating rates and the information radiation patterns for a focused beam, either in a running wave or a standing wave configuration. These results are relevant since in situations where the recoil heating rate dominates any other source of noise, the information radiation pattern is key to enable center-of-mass ground-state cooling via active feedback. We have shown that in this experimentally feasible configuration, high collection efficiency of the light carrying information about the center-of-mass degrees of freedom can be achieved for particles comparable and larger than the optical wavelength. This is possible not only in one degree of freedom as it happens for small particles, but for the three degrees of freedom, thereby allowing for simultaneously three-dimensional center-of-mass ground-state cooling.

Our work opens many research directions for further exploration, some of which we are currently investigating. (i)~As mentioned above, one could extend the results of this article to other shapes and degrees of freedom, such as rotation and internal acoustic vibrations. (ii)~Large particles support whispering gallery modes (WGMs) which can potentially enhance the coupling strength. These modes possess a high angular momentum and can be excited by  off-axis focused Gaussian beams (c.f.~\cite{ZambranaPuyalto2021}).  (iii)~While we have focused on silica particles, other dielectric particles have a much higher refractive index (e.g. silicon) which enhance optical resonances that could be exploited for optomechanics~\cite{Lepeshov2022}. It would be particularly interesting  to study and exploit the optical interaction between several dielectric particles beyond the point-dipole regime~\cite{Svak2021,GonzalezBallestero2021,Rieser2022,Vijayan2022,Rotter2023} and in the quantum regime. (iv)~The fact that dielectric particles beyond the point-dipole regime experience large recoil heating rates, comparable and even larger than the mechanical frequencies, hints at the possibility to enter the strong quantum optomechanical regime~\cite{Ranfagni2021,Sommer2021,Delic2020} in free space, which could have applications to generate quantum interfaces between light and center-of-mass motion without the use of optical resonators. (v)~Our quantum theory allows us to consider the injection of vacuum squeezing to modify the trade-off between imprecision and backacktion noise and obtain displacement sensitivities beyond the standard quantum limit, as recently discussed in~\cite{Ballestero2022} in the point-dipole approximation. This could be particularly interesting in the context of using micrometer-sized dielectric spheres for the search of new physics, such as dark matter~\cite{BlakemorePRA2019,MonteiroPRA2020,RademacherAOT2020,KawasakiRSI2020,MonteiroPRL2020,AfekPRD2021,BlakemorePRD2021,MooreQST2021,AfekPRL2022,PrielSciAdv2022,Carney2022}.

\begin{acknowledgments}

We acknowledge useful discussions with M. Aspelmeyer, M. Frimmer, D. H\"ummer, A. Militaru, L. Novotny, and M. Rossi. This research was supported by the European Union’s Horizon 2020 research and innovation programme under grant agreement No. [863132] (IQLev) and by the European Research Council (ERC) under the grant Agreement No. [951234] (Q-Xtreme ERC-2020-SyG).

\end{acknowledgments}

\clearpage

\bibliography{references}

\clearpage

\onecolumngrid
\appendix

\section{Derivation of explicit expressions}
\label{App:explicit}

In this section we derive the explicit expressions for the central quantities in the main article. First we derive in \appref{App:explicit_coupling} an explicit expression for the coupling rates. Then we derive in \appref{App:transition_amplitudes} an explicit expression for the Stokes and anti-Stokes transition amplitudes in the asymptotic limit. Finally, we combine these results in \appref{App:explicit_IRP_recoil} by deriving explicit expression for the information radiation patterns and recoil heating rates. 

\subsection{Derivation of the coupling rates} \label{App:explicit_coupling}
In this section we show how to derive an analytical expression for the coupling rates $g_{\kappa\kappa'\mu}$ starting from \eqnref{eq:gcouplings} in the main article, that is
\begin{equation}\label{eq:couplingRatesAppendix}
g_{\kappa\kappa'\mu}=r_{0 \mu} \frac{\sqrt{\omega_\kappa \omega_{\kappa'}}}{2} \lim_{r\to\infty}r^2\int \text{d}\Omega(\uv_r\cdot \uv_\mu) \left[\FF^*_{\kappa}(\rr)\cdot\FF_{\kappa'}(\rr)+\frac{1}{kk'}\nabla\times\FF^*_{\kappa}(\rr)\cdot\nabla\times\FF_{\kappa'}(\rr)\right].
\end{equation}
In particular, we show that the coupling rates for a dielectric sphere of arbitrary refractive index and size can be expressed as a sum over discrete angular momentum indices $l=1,2,\dots$, $m=-l,-l+1,\dots,l$, and polarization indices $p\in \lbrace \text{TE, TM}\rbrace$.

We proceed in three steps (see details below). First, we insert in \eqnref{eq:couplingRatesAppendix} the explicit far-field expression of the normalized scattering eigenmodes that can be written in terms of the vector spherical harmonic $\mathbf{X}_l^m(\theta,\phi)$. This leads to an integrand that consists of a sum of products of two vector spherical harmonics of different order $l$ and a $\mu-$dependent term, that is $\uv_r\cdot \uv_\mu$. As a second step we derive an analytical expression for these angular integrals. Finally, we use trigonometric identities to further simplify the analytical expression for the coupling rates $g_{\kappa\kappa'\mu}$. 

Let us now start by deriving the explicit far-field expression of the scattering eigenmodes $\mathbf{F}_\kappa(\rr)$ in terms of vector spherical harmonics. In~\cite{Maurer2021} we have derived an expression for the scattering eigenmodes in terms of the spherical eigenmodes in spherical coordinates, namely $\mathbf{F}_\kappa(\rr)=k^{-1}\sum_{lmp}\bar{c}_{lmg}^p(\theta_k,\phi_k)\mathbf{S}_{lmg}^p(k;\rr)$. The spherical eigenmodes $\mathbf{S}_{lmg}^p(k;\rr)$ are characterized by the discrete angular momentum indices $l$, $m$, polarization indices $p$, and can be mapped to the electromagnetic fields of electric and magnetic multipoles. The coefficients $\bar{c}_{lmg}^p$ can be shown to read
\begin{align}\label{eq:expansionCoefficients1}
    \bar{c}_{lmg}^\text{TE}(\theta_k,\phi_k)&\equiv \Theta(l-|m|)c_{lmg}^\text{TE}(\theta_k,\phi_k)=\im^{l}\Theta(l-|m|)\mathbf{X}_l^{m*}(\theta_k,\phi_k)\cdot \uv_g(\theta,\phi),\\\label{eq:expansionCoefficients2}
    \bar{c}_{lmg}^\text{TM}(\theta_k,\phi_k)&\equiv \Theta(l-|m|)c_{lmg}^\text{TM}(\theta_k,\phi_k)=\im^{l}\Theta(l-|m|)\mathbf{X}_l^{m*}(\theta_k,\phi_k)\cdot [\uv_r\times\uv_g(\theta,\phi)].
\end{align}
where $\mathbf{X}_l^m(\theta,\phi)$ denotes a vector spherical harmonic as defined in~\cite{Maurer2021, Hill1954}, $\Theta(x)$ denotes the Heaviside theta function with $\Theta(0)=1$, and the two transverse polarization vectors read $\uv_1(\theta,\phi)\equiv \im \uv_\phi(\theta,\phi) $ and $\uv_2(\theta,\phi)\equiv \uv_\theta(\theta,\phi)$. For notational convenience the sum runs over all $l$ and $m$, where the restriction to $m=-l,-l+1,\dots,l$ is taken care of by the expansion coefficients $\bar{c}_{lmg}^p(\theta_k,\phi_k)$ that vanish for $l< |m|$.

The spherical eigenmodes $\mathbf{S}_{lmg}^p(k;\rr)$ are separable in the radial and angular variables $r$, $\theta$, and $\phi$. As shown explicitly in~\cite{Maurer2021}, their dependence on the radial coordinate is given by the spherical Bessel and Hankel functions $j_l(x)$ and $h_l(x)$ of the first kind and order $l$. Their dependence on the angular coordinates is given by vector spherical harmonics. Using $\lim_{r \rightarrow \infty} j_{l}(kr)= (kr)^{-1}\sin(kr-l\pi/2)$ and $\lim_{r\rightarrow \infty} h_l(kr)=-\im (kr)^{-1} \exp[\im(kr-l\pi/2)]$ it  follows that 
\begin{align}\label{eq:ff1}
  \lim_{r\to\infty}  \mathbf{F}_\kappa(\rr)&= \frac{1}{kr}\sum_{l m}[X_{lmg}^\text{TE}(kr,\theta_k,\phi_k)\mathbf{X}_l^m(\theta,\phi)+ Y_{lmg}^\text{TM}(kr,\theta_k,\phi_k)\mathbf{Y}_l^m(\theta,\phi)],\\\label{eq:ff2}
  \lim_{r\to\infty}  \nabla\times\mathbf{F}_\kappa(\rr)&= \frac{\im k }{kr}\sum_{lm }[X_{lmg}^\text{TM}(kr,\theta_k,\phi_k)\mathbf{X}_l^m(\theta,\phi)- Y_{lmg}^\text{TE}(kr,\theta_k,\phi_k)\mathbf{Y}_l^m(\theta,\phi)].
\end{align}
The far-field expressions of the normalized scattering eigenmodes are written in terms of a linear combination of the vector spherical harmonics $\mathbf{X}_l^m(\theta,\phi)$, and $\mathbf{Y}_l^m(\theta,\phi)=-\im \uv_r\times \mathbf{X}_l^m(\theta,\phi)$. The radial functions $X_{lmg}^p(kr,\theta_k,\phi_k)$ and $Y_{lmg}^p(kr,\theta_k,\phi_k)=X_{lmg}^p(kr-\pi/2,\theta_k,\phi_k)$ read
\begin{align}
    X_{lmg}^p(kr,\theta_k,\phi_k) &= \sqrt{\frac{2}{\pi}}\bar{c}_{lmg}^p(\theta_k,\phi_k)[\sin(kr-l\pi/2)-(-\im)^l\sin \varphi_l^p\exp[\im (kr-\varphi_l^p)]].
\end{align}
The explicit expression for $\varphi_l^p$ in terms of the Lorenz-Mie coefficients depend on the sphere's radius $R$ and relative permittivity $\epsilon$ and can be found in~\cite{Maurer2021}. They can also be expressed in terms of the Mie coefficients in~\cite{Bohren2004}, namely $a_l\equiv \im \sin\varphi_l^\text{TM}\exp(-\im\varphi_l^\text{TM})$ and $b_l=\im \sin\varphi_l^\text{TE}\exp(-\im\varphi_l^\text{TE})$. Inserting \eqnref{eq:ff1} and \eqnref{eq:ff2} in \eqnref{eq:couplingRatesAppendix} and using $\omega_\kappa= ck$, we arrive at
\begin{multline}\label{eq:coupling_intermediate_step}
  g_{\kappa\kappa'\mu}=\frac{c r_{0\mu} }{2\sqrt{kk'}}\sum_{lm}\sum_{l'm'}\int \text{d}\Omega(\uv_r\cdot \uv_\mu)\\
  \times \lbrace [X_{lmg}^\text{TE}(kr,\theta_k,\phi_k)\mathbf{X}_l^m(\theta,\phi)+Y_{lmg}^\text{TM}(kr,\theta_k,\phi_k)\mathbf{Y}_l^m(\theta,\phi)]^* \cdot [X_{l'm'g'}^\text{TE}(k'r,\theta_{k'},\phi_{k'})\mathbf{X}_{l'}^{m'}(\theta,\phi)+Y_{l'm'g'}^\text{TM}(k'r,\theta_{k'},\phi_{k'})\mathbf{Y}_{l'}^{m'}(\theta,\phi)]\\
  +[X_{lmg}^\text{TM}(kr,\theta_k,\phi_k)\mathbf{X}_l^m(\theta,\phi)-Y_{lmg}^\text{TE}(kr,\theta_k,\phi_k)\mathbf{Y}_l^m(\theta,\phi)]^*\cdot [X_{l'm'g'}^\text{TM}(k'r,\theta_{k'},\phi_{k'})\mathbf{X}_{l'}^{m'}(\theta,\phi)-Y_{l'm'g'}^\text{TE}(k'r,\theta_{k'},\phi_{k'})\mathbf{Y}_{l'}^{m'}(\theta,\phi)]\rbrace,
\end{multline}

In order to further evaluate \eqnref{eq:coupling_intermediate_step} we need to find an expression for the angular integrals containing the product of two vector spherical harmonics and the $\mu-$dependent term $\uv_r\cdot \uv_\mu$, that is
\begin{align}
    I^\mu_{XX}=I^\mu_{YY}&\equiv \int_{\mathbb{S}^2}\text{d}
    \Omega(\uv_r\cdot\uv_\mu)\mathbf{X}_l^{m*}(\theta,\phi)\cdot \mathbf{X}_{l'}^{m'}(\theta,\phi),\\
     I^\mu_{XY}&\equiv \int_{\mathbb{S}^2}\text{d}
    \Omega(\uv_r\cdot\uv_\mu)\mathbf{X}_l^{m*}(\theta,\phi)\cdot \mathbf{Y}_{l'}^{m'}(\theta,\phi),\\
    I^\mu_{YX}&\equiv \int_{\mathbb{S}^2}\text{d}
    \Omega(\uv_r\cdot\uv_\mu)\mathbf{Y}_l^{m*}(\theta,\phi)\cdot \mathbf{X}_{l'}^{m'}(\theta,\phi),
\end{align}
where we have used that $[\mathbf{Y}_{l}^m(\theta,\phi)]^*\cdot \mathbf{Y}_{l'}^{m'}(\theta,\phi)=[\mathbf{X}_{l}^m(\theta,\phi)]^*\cdot \mathbf{X}_{l'}^{m'}(\theta,\phi)$. Using the results of~\cite{Neves2019} it is possible to derive an analytical expression for each of the above terms, namely
\begin{align}\notag
    I_{XX}^x&=\sqrt{\frac{l(l+2)(l-m'+1)(l-m')}{(2l+3)(2l+1)}}\frac{\delta_{l+1l'}\delta_{mm'+1}}{2l+2}-\sqrt{\frac{l'(l'+2)(l'+m'+2)(l'+m'+1)}{(2l'+3)(2l'+1)}}\frac{\delta_{ll'+1}\delta_{mm'+1}}{2l'+2}\\ \label{eq:Ix_XX}
    &+\sqrt{\frac{l'(l'+2)(l'-m+1)(l'-m)}{(2l'+3)(2l'+1)}}\frac{\delta_{ll'+1}\delta_{m+1m'}}{2l'+2}-\sqrt{\frac{l(l+2)(l+m+2)(l+m+1)}{(2l+3)(2l+1)}}\frac{\delta_{l+1l'}\delta_{m+1m'}}{2l+2},\\
    I_{XY}^x&=I_{YX}^x=-\frac{\sqrt{(l+m)(l-m+1)}}{2l(l+1)}\delta_{ll'}\delta_{mm'+1}-\frac{\sqrt{(l-m)(l+m+1)}}{2l(l+1)}\delta_{ll'}\delta_{m+1m'},\\
    \notag
    I_{XX}^y&=\sqrt{\frac{l(l+2)(l-m'+1)(l-m')}{(2l+3)(2l+1)}}\frac{\delta_{l+1l'}\delta_{mm'+1}}{2\im l+2 \im}-\sqrt{\frac{l'(l'+2)(l'+m'+2)(l'+m'+1)}{(2l'+3)(2l'+1)}}\frac{\delta_{ll'+1}\delta_{mm'+1}}{2 \im l'+2 \im}\\
    &-\sqrt{\frac{l'(l'+2)(l'-m+1)(l'-m)}{(2l'+3)(2l'+1)}}\frac{\delta_{ll'+1}\delta_{m+1m'}}{2\im l'+2\im}+\sqrt{\frac{l(l+2)(l+m+2)(l+m+1)}{(2l+3)(2l+1)}}\frac{\delta_{l+1l'}\delta_{m+1m'}}{2\im l+2\im },\\
    I_{YX}^y&=I_{XY}^y=-\frac{\sqrt{(l+m)(l-m+1)}}{2\im l(l+1)}\delta_{ll'}\delta_{mm'+1}+\frac{\sqrt{(l-m)(l+m+1)}}{2\im l(l+1)}\delta_{ll'}\delta_{m+1m'},\\
    I_{XX}^z&=\sqrt{\frac{l(l+2)(l+m'+1)(l-m'+1)}{(2l+3)(2l+1)}}\frac{\delta_{l+1l'}\delta_{mm'}}{l+1}+\sqrt{\frac{l'(l'+2)(l'+m+1)(l'-m+1)}{(2l'+3)(2l'+1)}}\frac{\delta_{ll'+1}\delta_{mm'}}{l'+1},\\
    I_{XY}^z&=-\frac{m}{l(l+1)}\delta_{ll'}\delta_{mm'}=I_{YX}^z.\label{eq:Iz_YX}
\end{align}

Note that all integrals contain a Kronecker delta in $l'$ and $m'$. Thus, inserting Eqs.(\ref{eq:Ix_XX})-(\ref{eq:Iz_YX}) in \eqnref{eq:coupling_intermediate_step} leads to an expression for the coupling rates $g_{\kappa\kappa'\mu}$ that is written as a \textit{single} sum over the discrete angular momentum and polarization indices. For each $\mu \in \lbrace x,y,z\rbrace$ the sum runs over terms that are all proportional to one of the two following combinations of products of the radial functions $X_{lmg}^p(kr,\theta_k,\phi_k)$ and $Y_{lmg}^p(kr,\theta_k,\phi_k)$, namely 
\begin{multline}\label{eq:radialFunction1}
	X_{lmg}^{p*}(kr,\theta_k,\phi_k)X^p_{l'm'g'}(k'r,\theta_{k'},\phi_{k'})+Y_{lmg}^{p*}(kr,\theta_{k},\phi_{k})Y^p_{l'm'g'}(k'r,\theta_{k'},\phi_{k'})=\\
	\frac{2}{\pi}\bar{c}^{p*}_{lmg}(\theta_{k},\phi_{k})\bar{c}^p_{l'm'g'}(\theta_{k'},\phi_{k'})\exp[\im(\varphi_l^p-\varphi^p_{l'})]\cos[(l-l')\pi/2+\varphi_l^p-\varphi_{l'}^p],
\end{multline}
\begin{multline}\label{eq:radialFunction2}
	X_{lmg}^{p*}(kr,\theta_{k},\phi_{k})Y^{p'}_{l'm'g'}(k'r,\theta_{k'},\phi_{k'})-Y_{lmg}^{p*}(kr,\theta_{k},\phi_{k})X^{p'}_{l'm'g'}(k'r,\theta_{k'},\phi_{k'})=\\
	\frac{2}{\pi}\bar{c}^{{p}*}_{lmg}(\theta_{k},\phi_{k})\bar{c}^{p'}_{l'm'g'}(\theta_{k'},\phi_{k'})\exp[\im(\varphi_l^{p}-\varphi^{p'}_{l'})]\sin[(l-l')\pi/2+\varphi_l^{p}-\varphi_{l'}^{p'}].
\end{multline}
The above equations immediately follow from standard trigonometric identities. As expected, this leads to an expression for the coupling rates that is independent on the radial variable $r$. 

Note that \eqnref{eq:radialFunction1} and \eqnref{eq:radialFunction2} do in principle depend on $k$ and $k'$ independently through $\varphi_l^p$ and $\varphi_{l'}^{p'}$. Here, we derive coupling rates for processes where photons at frequency $\omega_0$ interacting with center-of-mass phonons with frequencies $\Omega_\mu$ that are many orders of magnitude smaller $\Omega_\mu \ll \omega_0$. In light of this and in order to simplify the notation we do therefore approximate $k\simeq k'$. Combining all these results it immediately follows that the coupling rates can be compactly written as
\begin{equation}\label{eq:couplingratesANotation}
    g_{\kappa\kappa'\mu}=\frac{\im c r_{0\mu} }{2\pi \sqrt{kk'}}\sum_{lmp}\bar{c}^{p*}_{lmg}(\theta_{k},\phi_{k})a^{\mu p}_{lmg'}(\theta_{k'},\phi_{k'}).
\end{equation}
Here we have used the fact that the sum runs over all $l\in \mathds{N}^+$ and $m \in \mathds{Z}$ which allows for a simple shifting of the summation indices. The above representation is particularly advantageous for deriving an expression for the IRPs and the recoil heating rates  in \appref{App:explicit_IRP_recoil}. The explicit form of $a_{lmg}^{\mu p}(\theta_{k},\phi_{k})$ reads
\begin{align}\notag 
	a_{lmg}^{xp}(\theta_{k},\phi_{k})&=\sqrt{\frac{l(l+2)(l-m+2)(l-m+1)}{(2l+3)(2l+1)(2l+2)^2}}\bar{c}_{l+1m-1g}^pS^p_{l}-\sqrt{\frac{(l-1)(l+1)(l-m)(l-m-1)}{(2l+1)(2l-1)(2l)^2}}\bar{c}_{l-1m+1}^p(S^p_{l-1})^*\\\notag
	&-\sqrt{\frac{l(l+2)(l+m+2)(l+m+1)}{(2l+3)(2l+1)(2l+2)^2}}\bar{c}_{l+1m+1g}^pS^p_{l}+\sqrt{\frac{(l-1)(l+1)(l+m)(l+m-1)}{(2l+1)(2l-1)(2l)^2}}\bar{c}^p_{l-1m-1g}(S^p_{l-1})^*\\ \label{eq:a1}
	&-\frac{\sqrt{(l+m)(l-m+1)}}{2l(l+1)}\bar{c}_{lm-1g}^{\bar{p}}R_l^{\bar{p}}-\frac{\sqrt{(l+m+1)(l-m)}}{2l(l+1)}\bar{c}_{lm+1g}^{\bar{p}}R_l^{\bar{p}},\\\notag
	\im a_{lmg}^{yp}(\theta_{k},\phi_{k})&=\sqrt{\frac{l(l+2)(l-m+2)(l-m+1)}{(2l+3)(2l+1)(2l+2)^2}}\bar{c}_{l+1m-1g}^pS_{l}^p+\sqrt{\frac{(l-1)(l+1)(l-m)(l-m-1)}{(2l+1)(2l-1)(2l)^2}}\bar{c}_{l-1m+1}^p(S_{l-1}^p)^*\\\notag
	&+\sqrt{\frac{l(l+2)(l+m+2)(l+m+1)}{(2l+3)(2l+1)(2l+2)^2}}\bar{c}_{l+1m+1g}^pS_{l}^p+\sqrt{\frac{(l-1)(l+1)(l+m)(l+m-1)}{(2l+1)(2l-1)(2l)^2}}\bar{c}^p_{l-1m-1g}(S_{l-1}^p)^*\\ \label{eq:a2}
	&-\frac{\sqrt{(l+m)(l-m+1)}}{2l(l+1)}\bar{c}_{lm-1g}^{\bar{p}}R_l^{\bar{p}}+\frac{\sqrt{(l+m+1)(l-m)}}{2l(l+1)}\bar{c}_{lm+1g}^{\bar{p}}R_l^{\bar{p}},\\\notag
	a_{lmg}^{zp}(\theta_{k},\phi_{k})&= \sqrt{\frac{l(l+2)[(l+1)^2-m^2]}{(2l+3)(2l+1)(l+1)^2}}\bar{c}_{l+1mg}^pS_{l}^p-\sqrt{\frac{(l^2-1)(l^2-m^2)}{(4l^2-1)l^2}}\bar{c}_{l-1mg}^p(S_{l-1}^p)^*\\ \label{eq:a3}
	&-\frac{m}{l(l+1)}\bar{c}^{\bar{p}}_{lmg}R_l^{\bar{p}},
\end{align}
where we have defined $\overline{\text{TE}}\equiv \text{TM}$, $\overline{\text{TM}}\equiv \text{TE}$, and
\begin{align}
    R_l^\text{TM}=-(R_l^\text{TE})^*&\equiv 1-\exp[-2\im(\varphi_{l}^\text{TM}-\varphi_l^\text{TE})],\\
    S_{l}^p&\equiv 1-\exp[-2\im(\varphi_{l+1}^p-\varphi_{l}^p)].
\end{align}
For better readability we have chosen to not specify explicitly that $\bar{c}_{lmg}^p$ depends on $(\theta_{k'},\phi_{k'})$ in the above expressions. This concludes the derivation of the coupling rates for spheres of arbitrary refractive index and size. 

\subsubsection{Small-particle limit}

Finally, let us explain how to derive the small particle limit of the coupling rates, that is \eqnref{eq:gcouplings_small_particle} in the main article. The small particle limit $(\sqrt{\epsilon}kR\ll 1)$ is obtained by realizing that, through $R_l^p$ and $S_l^p$, the contribution of the individual terms of different $l\in\mathds{N}^+$ in~\eqnref{eq:couplingratesANotation} to the total sum decreases exponentially with increasing $l$. For $q\equiv kR\to 0$ we have $R_l^p=\mathcal{O}(q^{2l+1})$, $S_l^\text{TE}=\mathcal{O}(q^{2l+3})$, and $S_l^\text{TM}=\mathcal{O}(q^{2l+1})$. The leading order terms read 
\begin{align}
    R_1^\text{TM}\simeq R_1^\text{TE}\simeq -S_1^\text{TM}&\simeq -\frac{4\im}{3}\frac{\epsilon-1}{\epsilon+2}(kR)^3=-\frac{\im \alpha k^3}{3\pi \epsilon_0},
\end{align}
where $\alpha = 3\epsilon_0 V(\epsilon-1)/(\epsilon+2)$ is the polarizability of the dielectric sphere and $V$ the volume of the sphere. Retaining only those terms~\eqnref{eq:couplingratesANotation} directly leads to
\begin{equation} 
	 g_{\kappa\kappa'\mu}\simeq \frac{\im r_{0\mu} \alpha}{\epsilon_0} \frac{c k k'  }{2 (2\pi)^3}(\uv_{g}^*\cdot \uv_{g'})(\uv_k-\uv_{k'})\cdot \uv_{\mu},
\end{equation}
where $\uv_k$ denotes the unit vector parallel to the wave vector $\kk$. This expression agrees with the heuristically derived expressions for the coupling rates, see \secref{sec:fundamentalHamiltonian}. 

\subsection{Derivation of the transition amplitudes}\label{App:transition_amplitudes}
In this section we consider Stokes and anti-Stokes scattering processes as shown in  \figref{fig1} and derive the corresponding transition amplitudes in first-order perturbation theory in the asymptotic limit (c.f. \eqnref{eq:asymptotic_TA}). These processes describe the scattering of a single photon from a coherently populated tweezer mode into a plane-wave mode $\kappa$ with annihilation and creation operators $\hat{b}_\kappa, \hat{b}_\kappa^\dagger$~\cite{Maurer2021} (not excluding the tweezer mode) by generating or absorbing a center-of-mass phonon through the dynamics generated by the fundamental Hamiltonian $\hat{H}=\hat{H}_0+\hat{H}_\text{int}$ in~\eqnref{eq:QEDHamiltonianBosons}. 

The transition amplitudes for this processes read $\tau^p_{\kappa \mu}(t,t')\equiv \bra{\Psi^p_\text{out}}\hat{U}(t,t')\ket{\Psi_\text{in}}$, with the input state $\ket{\Psi_\text{in}}\equiv \ket{n_x,n_y,n_z} \otimes \ket{\Psi_\text{cl}}$ ($n_\mu >0$), the output states $\ket{\Psi^\text{S}_\text{out}}\equiv \hat{b}_\kappa^\dagger \hat{b}_\mu^\dagger \ket{\Psi_\text{in}}$ and  $\ket{\Psi^\text{aS}_\text{out}}\equiv \hat{b}_\kappa^\dagger \hat{b}_\mu \ket{\Psi_\text{in}}$ describing Stokes ($p=\text{S}$) and anti-Stokes ($p=\text{aS}$) processes respectively, the time-evolution operator in the Schr\"odinger picture $\hat{U}(t,t')$, and $\ket{\Psi_\text{cl}}=\mathcal{D}(\alpha_{\kappa})\ket{0}_\text{em}$ as defined in \eqnref{eq:Eclassical} with the coherent and monochromatic amplitude $\alpha_{\kappa}$ with frequency $\omega_0$. We rewrite the transition amplitude $\tau^p_{\kappa\mu}$ with $p\in \lbrace \text{S, aS}\rbrace$  as
\begin{align}
    \tau^\text{S}_{\kappa\mu} &= \exp(-\im \Omega_\mu t)\bra{\Psi_\text{in}}\exp(-\im \hat{H_0}t/\hbar) \hat{b}_\kappa(t) \hat{b}_\mu \hat{U}_\text{int}(t,t')\exp(\im \hat{H_0}t'/\hbar) \ket{\Psi_\text{in}},\\
    \tau^\text{aS}_{\kappa\mu} &=\exp(\im \Omega_\mu t)\bra{\Psi_\text{in}}\exp(-\im \hat{H_0}t/\hbar) \hat{b}_\kappa(t) \hat{b}^\dagger_\mu \hat{U}_\text{int}(t,t')\exp(\im \hat{H_0}t'/\hbar) \ket{\Psi_\text{in}},
\end{align}
where $\hat{U}_\text{int}(t,t')\equiv \exp(\im\hat{H_0}t/\hbar)\hat{U}(t,t')\exp(-\im \hat{H_0}t'/\hbar)$ denotes the time-evolution operator and $\hat{b}_\kappa(t)\equiv \exp(\im\hat{H_0}t/\hbar) \hat{b}_\kappa \exp(-\im\hat{H_0}t/\hbar)$ the plane-wave annihilation operator in the interaction picture.

In the regime where the coupling rates $g_{\kappa\kappa'\mu}$ are smaller than the trap frequencies $\Omega_\mu$ we expand $\hat{U}_\text{int}(t,t')$ up to first order in time-dependent perturbation theory and set $t,t'=\pm T/2$, namely
\begin{align}
\hat{U}_\text{int}(T/2,-T/2)&\simeq 1 -\frac{\im}{\hbar} \int_{-T/2}^{T/2} ds \exp(\im \hat{H}_0 s) \hat{H}_\text{int} \exp(-\im \hat{H}_0 s)\\
&=1-2\im \pi  \sum_{\kappa\kappa'\mu} g_{\kappa\kappa'\mu}\hat{a}_\kappa^\dagger \hat{a}_{\kappa'}[\hat{b}_\mu \delta_T(\omega_\kappa-\omega_{\kappa'}-\Omega_\mu)+ \hat{b}_\mu^\dagger \delta_T(\omega_\kappa-\omega_{\kappa'}+\Omega_\mu)],
\end{align}
where $\delta_T(\omega)\equiv (2\pi)^{-1}\int_{-T/2}^{T/2} ds \exp(\im \omega s) $ denotes a nascent delta function. This leads to
\begin{multline}
    \tau^\text{S}_{\kappa\mu}\simeq-2\im \pi(n_\mu +1) \exp(-\im \Omega_\mu T/2-\im \textstyle\sum_\nu n_\nu \Omega_\nu T)\displaystyle\sum_{\kappa_1\kappa_2}g_{\kappa_1\kappa_2\mu}\\
    \times\bra{\Psi_\text{cl}} \exp[-\im \hat{H_0}T/(2\hbar)]\hat{b}_\kappa(T/2) \hat{a}_{\kappa_1}^\dagger \hat{a}_{\kappa_2} \exp[-\im \hat{H_0}T/(2\hbar)]\ket{\Psi_\text{cl}}\delta_T(\omega_{\kappa_1}-\omega_{\kappa_2}+\Omega_\mu),\\
\end{multline}
for the Stokes process, and
\begin{multline}
    \tau^\text{aS}_{\kappa\mu}\simeq-2\im \pi n_\mu \exp(\im \Omega_\mu T/2-\im  \textstyle\sum_\nu n_\nu \Omega_\nu T)\displaystyle\sum_{\kappa_1\kappa_2}g_{\kappa_1\kappa_2\mu}\\
    \times\bra{\Psi_\text{cl}} \exp[-\im \hat{H_0}T/(2\hbar)]\hat{b}_\kappa(T/2) \hat{a}_{\kappa_1}^\dagger \hat{a}_{\kappa_2} \exp[-\im \hat{H_0}T/(2\hbar)]\ket{\Psi_\text{cl}}\delta_T(\omega_{\kappa_1}-\omega_{\kappa_2}-\Omega_\mu),\\
\end{multline}
for the anti-Stokes process, where we have used $\hat{b}_\mu \ket{n_\mu} = \sqrt{n_\mu}\ket{n_\mu-1}$ and $\hat{b}^\dagger_\mu \ket{n_\mu} = \sqrt{n_\mu+1}\ket{n_\mu+1}$. To further simplify the expression we use that $\mathcal{D}^\dagger(\alpha_{\kappa})\mathcal{D}(\alpha_{\kappa})= 1$, $\mathcal{D}^\dagger(\alpha_{\kappa}) a_\kappa \mathcal{D}(\alpha_{\kappa}) = \hat{a}_\kappa + \alpha_{\kappa}$, and $\hat{a}_\kappa \ket{0}_\text{em} = 0$. This leads to
\begin{multline}
    \sum_{\kappa_1\kappa_2}g_{\kappa_1\kappa_2\mu}\bra{\Psi_\text{cl}} \exp[-\im \hat{H_0}T/(2\hbar)]\hat{b}_\kappa(T/2) \hat{a}_{\kappa_1}^\dagger \hat{a}_{\kappa_2} \exp[-\im \hat{H_0}T/(2\hbar)]\ket{\Psi_\text{cl}}\delta_T(\omega_{\kappa_1}-\omega_{\kappa_2}\pm \Omega_\mu)\\
    =\exp(-\im \omega_0 T/2)\sum_{\kappa_1}G_{\kappa_1\mu}\bra{\Psi_\text{cl}} \exp[-\im \hat{H_0}T/(2\hbar)]\hat{b}_\kappa(T/2) \hat{a}_{\kappa_1}^\dagger  \exp[-\im \hat{H_0}T/(2\hbar)]\ket{\Psi_\text{cl}}\delta_T(\omega_{\kappa_1}-\omega_{0}\pm \Omega_\mu),
\end{multline}
where we have defined $G_{\kappa\mu }\equiv \sum_{\kappa'}\alpha_{\kappa'}g_{\kappa\kappa'\mu}$. Using $[a_{\kappa},a_{\kappa'}^\dagger]=\delta_{\kappa\kappa'}$, $\lim_{T\to\infty} [\hat{b}_\kappa(T)\exp(\im \omega_\kappa T)] =   \sum_{\kappa'}S_{\kappa\kappa'} \hat{a}_{\kappa'}$~\cite{Maurer2021}, and $\Omega_\mu>0$ we finally arrive at
\begin{align}\label{eq:transitionAmplitudeS}
    \tau^\text{S}_{\kappa\mu}=&-2\im \pi (n_\mu+1) \exp[-\im T(\omega_0+\textstyle\sum_\nu n_\nu \Omega_\nu)]\sum_{\kappa'}S_{\kappa\kappa'}G_{\kappa' \mu} \delta_T(\omega_{\kappa}-\omega_0+\Omega_\mu)  \bra{\Psi_\text{cl}}\exp(-\im \hat{H_0}T/\hbar)\ket{\Psi_\text{cl}},\\
    \label{eq:transitionAmplitudeaS}
    \tau^\text{aS}_{\kappa\mu}=&-2\im \pi n_\mu\exp[-\im T(\omega_0+\textstyle\sum_\nu n_\nu \Omega_\nu)]\sum_{\kappa'}S_{\kappa\kappa'}G_{\kappa' \mu} \delta_T(\omega_{\kappa}-\omega_0-\Omega_\mu) \bra{\Psi_\text{cl}}\exp(-\im \hat{H_0}T/\hbar)\ket{\Psi_\text{cl}}.
\end{align}

\subsection{Derivation of the IRPs and recoil heating rates} \label{App:explicit_IRP_recoil}

In this section we combine the results of the previous two sections in order to derive explicit expressions for the recoil heating rates~\eqnref{eq:recoilheatinggeneral} and the information radiation patterns~\eqnref{eq:informationRadiationPattern} associated to the center-of-mass motion along the $\mu$-axis.

Using the transition amplitudes $\mathcal{\tau}_{\kappa \mu}^p$ in \eqnref{eq:transitionAmplitudeS} and \eqnref{eq:transitionAmplitudeaS} we can derive the transition probability rates in the asymptotic limit~\cite{Cohen2004/2}, namely
\begin{align}
  \lim_{T\to \infty } \frac{\sum_\kappa|\tau_{\kappa\mu}^\text{S}|^2}{T}&=(n_\mu +1)\Gamma_\mu^+ ,\\
  \lim_{T\to \infty } \frac{\sum_\kappa|\tau_{\kappa\mu}^\text{aS}|^2}{T}&=n_\mu\Gamma_\mu^-,
\end{align}
for Stokes $(p=\text{S})$ and anti-Stokes $(p=\text{aS})$ processes respectively, and $\Gamma^{\pm}_\mu  \equiv 2\pi \sum_{\kappa} |\sum_{\kappa'} S_{\kappa\kappa'}G_{\kappa'\mu}|^2 \delta(\omega_\kappa - \omega_0 \pm \Omega_\mu)$. For coupling rates that are sufficiently broadband, i.e. $|\Gamma_\mu^+-\Gamma_\mu^-| \ll \Gamma_\mu^+ + \Gamma_\mu^-$ we define a single transition rate $\Gamma_\mu  = 2\pi \sum_{\kappa} |\sum_{\kappa'} S_{\kappa\kappa'}G_{\kappa'\mu}|^2 \delta(\omega_\kappa - \omega_0)$ which can be shown to correspond to the phonon heating rate due to laser recoil.

Inserting the expression for the linearized coupling rate $G_{\kappa\mu}$ we arrive at \begin{equation}
	\Gamma_\mu = 2\pi \sum_{\kappa} \left|\sum_{\kappa'\kappa''}S_{\kappa\kappa'}\alpha_{\kappa''}g_{\kappa'\kappa''\mu}\right|^2 \delta(\omega_\kappa - \omega_0),
\end{equation}
where $\alpha_\kappa$ is the coherent amplitude of the monochromatic coherent state $\ket{\psi_\text{cl}}$~(\eqnref{eq:Eclassical}) with frequency $\omega_\kappa=\omega_0$. Inserting~\eqnref{eq:couplingratesANotation} we arrive at
\begin{align}\label{eq:recoilHeatingnotSimplified}
    \Gamma_\mu=\frac{c r^2_{0\mu}}{2\pi}\sum_{lmp}\sum_{l'm'p'}\left[\sum_g\int\text{d}\Omega_k \bar{c}_{lmg}^{p*}(\theta_{k},\phi_{k})\bar{c}_{l'm'g}^{p'} (\theta_{k},\phi_{k})\right]A_{lm}^{\mu p}(A_{l'm'}^{\mu p'})^*\exp[-2\im (\varphi_{l}^p-\varphi_{l'}^{p'})],
\end{align}
where we have defined $A_{lm}^{\mu p}\equiv \sum_\kappa \alpha_\kappa a_{lmg}^{\mu p}(\theta_{k},\phi_{k})$. We can greatly simplify the above expression by evaluating the expression in square brackets. Inserting Eqs.(\ref{eq:expansionCoefficients1})-(\ref{eq:expansionCoefficients2}) we have for e.g. $p=p'=\text{TE}$ that
\begin{align}\notag
    &\sum_g\int\text{d}\Omega_k \bar{c}_{lmg}^{\text{TE}*}(\theta_{k},\phi_{k})\bar{c}_{l'm'g}^{\text{TE}'}(\theta_{k},\phi_{k})\\
    &=(-\im)^l \im^{l'}\Theta(l-|m|)\Theta(l'-|m'|)\int\text{d}\Omega_k \mathbf{X}_{l}^m(\theta_k,\phi_k)\cdot \sum_g \uv^*_g(\theta,\phi)\otimes \uv_g(\theta,\phi)\cdot \mathbf{X}_{l'}^{m'*}(\theta_k,\phi_k)\\
    &=(-\im)^l \im^{l'}\Theta(l-|m|)\Theta(l'-|m'|)\int\text{d}\Omega_k \mathbf{X}_{l}^m(\theta_k,\phi_k)\cdot (\mathds{1}-\uv_r\otimes\uv_r)\cdot \mathbf{X}_{l'}^{m'*}(\theta_k,\phi_k)\\
    &=(-\im)^l \im^{l'}\Theta(l-|m|)\Theta(l'-|m'|)\int\text{d}\Omega_k \mathbf{X}_{l}^m(\theta_k,\phi_k)\cdot \mathbf{X}_{l'}^{m'*}(\theta_k,\phi_k)=\Theta(l-|m|)\delta_{ll'}\delta_{mm'}.
\end{align}
To arrive to the third line we use the completeness relation of the unit vectors, where $\otimes$ denotes the dyadic product and to arrive to the fourth line we used the fact that $\mathbf{X}_l^m(\theta,\phi)$ has a vanishing radial component and finally in the last step we have used the orthogonality condition of the vector spherical harmonics. Analogous derivations for all remaining combinations of $p,p'\in \lbrace \text{TE,TM}\rbrace$ lead to 
\begin{equation}
\sum_g\int\text{d}\Omega_k \bar{c}_{lmg}^{p*}(\theta_{k},\phi_{k})\bar{c}_{l'm'g}^{p'} (\theta_{k},\phi_{k})=\Theta(l-|m|)\delta_{ll'}\delta_{mm'}\delta_{pp'}
\end{equation}
and hence
\begin{equation} \label{eq:recoil_explicit}
    \Gamma_\mu= \frac{c r^2_{0\mu}}{2\pi}\sum_{lmp}\Theta(l-|m|)\left| A_{lm}^{\mu p}\right|^2,
\end{equation}
where $A_{lm}^{\mu p}$ depends, through $\alpha_\kappa$, on the particular state of the electromagnetic field. 

The IRP $\mathcal{I}^p_\mu (\theta_k,\phi_k)$ is defined as the normalized angular distribution of the transition probability rate in the asymptotic limit, namely
\be 
    \mathcal{I}^p_\mu(\theta_k,\phi_k)\equiv  \frac{\lim_{T\to \infty } \int_0^\infty dk k^2 \sum_g |\tau^p_{\kappa\mu}|^2}{\lim_{T\to \infty } \sum_\kappa |\tau^p_{\kappa\mu}|^2},
\ee
with $\int d\Omega_k \mathcal{I}_\mu(\theta_k,\phi_k) = 1.$ Within broadband coupling regime, one obtains $\mathcal{I}_\mu(\theta_k,\phi_k)\equiv \mathcal{I}^\text{S}_\mu(\theta_k,\phi_k)= \mathcal{I}_\mu^\text{aS}(\theta_k,\phi_k)\equiv$. Inserting \eqnref{eq:transitionAmplitudeS} first and then \eqnref{eq:couplingratesANotation} it follows that the IRP reads
\begin{equation}\label{eq:IRPexplicit}
    \mathcal{I}_\mu(\theta_k,\phi_k)=\frac{\sum_{g}\int_0^\infty dk k^2 |\sum_{\kappa'}S_{\kappa\kappa'}G_{\kappa'\mu}|^2 \delta(\omega_\kappa - \omega_0)}{\sum_{\kappa} |G_{\kappa\mu}|^2 \delta(\omega_\kappa - \omega_0)}= \frac{\sum_g \left|\sum_{lmp}\bar{c}_{lmg}^{p*}(\theta_{k},\phi_{k})A_{lm}^{\mu p} \exp(-2\im\varphi_l^p)\right|^2}{\sum_{lmp}\Theta(l-|m|)\left| A_{lm}^{\mu p}\right|^2}.
\end{equation}

\section{Focused Gaussian beam as a linear combination of plane waves} \label{App:gaussian}

Both central quantities of this article, namely the recoil heating rate in~\eqnref{eq:recoilheatinggeneral} and the information radiation pattern in~\eqnref{eq:informationRadiationPattern} crucially depend on the coherent amplitude $\alpha_\kappa$ that specifies the particular state of the electromagnetic field. In this section we derive an explicit expression for $\alpha_\kappa$ for a single focused Gaussian beam and a standing wave configuring of two counter-propagating focused Gaussian beams with an arbitrary phase shift. 

Let us start by deriving a general expression for the electric field in the focal region of an aplanatic lens, i.e. a lens designed to minimize chromatic aberrations. We can then use this expression to infer $\alpha_\kappa$ through~\eqnref{eq:classicalField}. We consider a monochromatic incoming field $\mathbf{E}_\text{in}(\rr,t)\equiv \mathbf{E}_\text{in}(\rr)\exp(-\im \omega_0 t)+\text{c.c.}$ of frequency $\omega_0$ and wavelength $\lambda_0 =  2\pi c/\omega_0$. The aplanatic lens is characterized by its focal length $f$ and numerical aperture $\text{NA}=\sin\theta_\text{NA}$. Note that the evanescent components of the incident electric field make a negligible contribution to the focal field given that $f\gg \lambda_0$. Under this approximation, one can show shown that the electric field in the focal region reads~\cite{richards1959,novotny_hecht_2012}
\begin{align}\label{eq:focalField}
	\mathbf{E}(\rr)&=\frac{k_0 f}{2\pi \im }\int_{\mathcal{D}_\pm} \text{d}
\Omega_k \mathbf{E}^\pm_\infty(f,\theta_k,\phi_k)\exp[\im k r \cos\theta_k \cos\theta +\im k r \sin \theta_k\sin\theta \cos(\phi_k-\phi)],
\end{align}
where
\begin{align}\notag
	\mathbf{E}^\pm_\infty(f,\theta_k,\phi_k)&=\sqrt{|\cos\theta_k|}[\mathbf{E}_\text{in}(f,\theta_k,\phi_k)\cdot \uv_{\phi}(\theta_k,\phi_k)]\uv_{\phi}(\theta_k,\phi_k)\\ \label{eq:farField}
	&\pm\sqrt{|\cos\theta_k|}[\mathbf{E}_\text{in}(f,\theta_k,\phi_k)\cdot \uv_{\rho}(\theta_k,\phi_k)]\uv_{\theta}(\theta_k,\phi_k),
\end{align}
denotes the far-field angular spectrum of the refracted for an incoming field propagating along the positive $(+)$ or negative $(-)$ $z$-axis, and $\uv_\phi$ and $\uv_\rho$ denote the unit vectors of the corresponding cylindrical coordinate system. The corresponding integration domains depend on the numerical aperture and read $\mathcal{D}_+=\lbrace \theta_k, \phi_k \in \mathds{R}\,|\,0 \leq \theta_k \leq \theta_\text{NA},\,0\leq \phi_k < 2 \pi\rbrace$ and $\mathcal{D}_-=\lbrace \theta_k,\phi_k \in \mathds{R}\,|\,\pi-\theta_\text{NA}\leq \theta_k \leq \pi,\,0\leq \phi_k < 2 \pi \rbrace$. Note that \eqnref{eq:farField} is derived on the premise that, upon refraction by the lens, the energy of the field is conserved and that the angle between the polarization and propagation of the field remains constant. 

Inserting \eqnref{eq:farField} in \eqnref{eq:focalField}, we immediately see that we can write $\mathbf{E}(\rr)= \im \sum_\kappa \sqrt{\hbar\omega_k/(2\epsilon_0)}\alpha_\kappa \mathbf{G}_\kappa(\rr)$ with the normalized plane-wave mode $\mathbf{G}_\kappa(\rr)=\exp(\im \kk \cdot \rr)\uv_g/(2\pi)^{3/2}$. Thus, the focused field $\mathbf{E}(\rr)$ is given by a linear combination of plane waves with coefficients that depend on the angular spectrum and the propagation direction $(\pm)$ of the incoming beam, namely
\begin{align}
    \alpha_1^\pm(k,\theta_k,\phi_k)&= -\im \sqrt{\frac{4\pi \epsilon_0}{\hbar \omega_0}} k_0 f \sqrt{|\cos\theta_k|}[\mathbf{E}_\text{in}(f,\theta_k,\phi_k)\cdot \uv_{\phi}(\theta_k,\phi_k)]\frac{\delta(k-k_0)}{k_0^2},\\
    \alpha_2^\pm(k,\theta_k,\phi_k)&=\pm \sqrt{\frac{4\pi \epsilon_0}{\hbar \omega_0}} k_0 f \sqrt{|\cos\theta_k|}[\mathbf{E}_\text{in}(f,\theta_k,\phi_k)\cdot \uv_{\rho}(\theta_k,\phi_k)]\frac{\delta(k-k_0)}{k_0^2},
\end{align}
where $k_0 \equiv \omega_0/c$. Note that in the presence of a dielectric sphere of arbitrary refractive index and size, the focused fields are obtained by simply replacing the plane-wave modes $\mathbf{G}_\kappa(\rr)$ by the scattering eigenmodes $\mathbf{F}_\kappa(\rr)$~\cite{Maurer2021}, that is $\mathbf{E}(\rr)= \im \sum_\kappa  \sqrt{\hbar\omega_k/(2\epsilon_0)}\alpha_\kappa \mathbf{F}_\kappa(\rr)$.

Having derived a general expression for the focal field let us now consider a single incoming $x-$polarized paraxial Gaussian beam of waist $w_0$ propagating along the positive/negative $z-$axis, that is $\mathbf{E}_\text{in}(f,\theta_k,\phi_k)=E_0\exp(-f^2\sin\theta^2_k/w_0^2)\uv_x$. This leads to 
\begin{align}\label{eq:alpha1}
    \alpha_1^{\pm}(k,\theta_k,\phi_k)&= \sqrt{\frac{4\pi \epsilon_0}{\hbar \omega_0}}k_0fE_0 \sqrt{|\cos\theta_k|}\exp(-f^2\sin^2\theta_k/w_0^2)\im \sin\phi_k\frac{\delta(k-k_0)}{k_0^2},\\
    \label{eq:alpha2}
    \alpha_2^{\pm}(k,\theta_k,\phi_k)&=\pm \sqrt{\frac{4\pi \epsilon_0}{\hbar \omega_0}}k_0fE_0 \sqrt{|\cos\theta_k|}\exp(-f^2\sin^2\theta_k/w_0^2)\cos\phi_k\frac{\delta(k-k_0)}{k_0^2}. 
\end{align}
The coherent amplitude enters in the expressions for the recoil heating rate and IRP through $A_{lm\pm}^{\mu p}=\sum_\kappa \alpha^\pm_\kappa a_{lmg}^{\mu p}(\theta_k,\phi_k)$. The Dirac delta in $k-k_0$ renders the integration over $k$ trivial and we only need to evaluate the sum over the polarizations $g$ and the integral over the angles $\theta_k,\phi_k$. One can see in Eqs. (\ref{eq:a1})-(\ref{eq:a3}) that the only quantity that depends on these variables is $\bar{c}_{lmg}^p(\theta_k,\phi_k)$. In order to further simplify the computation of the recoil heating rate and IRP, let us therefore derive an explicit expression for $C_{lm \pm}^p\equiv \sum_\kappa \alpha^\pm_\kappa c_{lmg}^p(\theta_k,\phi_k)$. With this quantity we immediately obtain an expressions for $A_{lm \pm}^{\mu p}$ by replacing $\bar{c}_{lmg}^p(\theta_k,\phi_k)$ by $C_{lm \pm}^{ p}$ in Eqs. (\ref{eq:a1})-(\ref{eq:a3}). Inserting Eqs. (\ref{eq:expansionCoefficients1})-(\ref{eq:expansionCoefficients2}) in $C_{lm \pm}^p\equiv \sum_\kappa \alpha^\pm_\kappa c_{lmg}^p(\theta_k,\phi_k)$ we have
\begin{align}
	C_{lm+}^\text{TE}&= \im^l \Theta(l-|m|)k_0fE_0 \sqrt{\frac{4\pi \epsilon_0}{\hbar \omega_0}} \int_{\mathcal{D}_+}\text{d}\Omega_k\sqrt{|\cos\theta_k|}\exp(-f^2\sin^2\theta_k/w_0^2)\mathbf{X}_l^{m*}(\theta_k,\phi_k)\cdot( -\sin\phi_k \uv_\phi+\cos\phi_k \uv_\theta),\\
	C_{lm+}^\text{TM}&= \im^l \Theta(l-|m|)k_0fE_0  \sqrt{\frac{4\pi \epsilon_0}{\hbar \omega_0}} \int_{\mathcal{D}_+}\text{d}\Omega_k\sqrt{|\cos\theta_k|}\exp(-f^2\sin^2\theta_k/w_0^2)\mathbf{X}_l^{m*}(\theta_k,\phi_k)\cdot(\sin\phi_k \uv_\theta+\cos\phi_k \uv_\phi).
\end{align}
Integration over $\phi_k$ can be performed analytically and leads to
\begin{equation}
    C_{lm+}^p= \im^l k_0fE_0 \sqrt{\frac{4\pi^3 \epsilon_0}{\hbar \omega_0}} (\mathbf{Z}_{lm}\cdot  \mathbf{u}^p_{+}\delta_{m1}+\mathbf{Z}_{lm}\cdot \mathbf{u}_{+}^{*p}\delta_{m-1}),
\end{equation}
where we have defined the two vectors $\mathbf{u}^\text{TE}_+\equiv \uv_\theta+\im \uv_\phi$, $\mathbf{u}^\text{TM}_+\equiv \uv_\phi-\im \uv_\theta$, and
\begin{align}
 	\mathbf{Z}_{lm}\equiv \Theta(l-|m|)\int_{0}^{\theta_\text{NA}} \text{d}\theta \sin\theta \sqrt{|\cos\theta|} \exp(-f^2 \sin^2\theta/w^2)\mathbf{X}_l^{m*}(\theta,0).
 \end{align}
Analogously, using  $\mathbf{X}_l^{\pm 1}(\pi-\theta_k,0)=(-1)^{l+1}[\mathbf{X}_l^{\pm 1}(\theta_k,0)]^*$, one obtains
\begin{equation}
    C_{lm-}^p= - (-\im)^l k_0fE_0 \sqrt{\frac{4\pi^3 \epsilon_0}{\hbar \omega_0}} (\mathbf{Z}^*_{lm}\cdot  \mathbf{u}_-^p\delta_{m1}+\mathbf{Z}^*_{lm}\cdot \mathbf{u}_-^{*p}\delta_{m-1}),
\end{equation}
where we have defined $\mathbf{u}^\text{TE}_-\equiv -\uv_\theta+\im \uv_\phi$, and $\mathbf{u}_-^\text{TM}\equiv -\uv_\phi-\im \uv_\theta$. We have now derived an expression for $C_{lm\pm}^p$ in the case of focused Gaussian beams propagating along the positive/negative $(+/-)$ z-axis. For the standing-wave configuration we combine both beams with a relative phase $\Phi$ which leads to
\begin{equation}
    \tilde{C}_{lm}^p(\Phi)\equiv C_{lm+}^{ p}+\exp(\im \Phi)C_{lm-}^{ p},
\end{equation}
where the intensity at the focus scales as $\cos^2(\Phi/2)$ with the relative phase.

\subsection{Characterization of the focused field}\label{app:charactFocusField}
Note that we can relate the electric field amplitude $E_0$ to the power $P$ that passes through a single lens, that is
\begin{align}\label{eq:powerOnLens}
    P&= 2\mu_0^{-1}f^2\int_\mathcal{D} d\Omega_k [ \mathbf{E}_\infty(f,\theta_k,\phi_k) \times \mathbf{B}^*_\infty(f,\theta_k,\phi_k) ] \cdot \uv_r(\theta_k,\phi_k)=\epsilon_0 c E_0^2\pi w^2[1-\exp(-2f^2\sin^2\theta_\text{NA}/w_0^2)],
\end{align}
where $\mu_0$ denotes the vacuum permeability and $\mathbf{B}(\rr,t)=\mathbf{B}(\rr)\exp(-\im \omega_0t)+\text{c.c}$ denotes the magnetic field. In order to achieve maximum focusing one needs to overfill the lens, that is $f_0\gg 1$ with filling factor $f_0\equiv w_0/(f\sin\theta_\text{NA})$. In this case we can approximate $P\simeq 2 \pi \epsilon_0 c (fE_0)^2\sin^2\theta_{\text{NA}}$. Throughout the article we consider a filling factor $f_0 = 10$.

Let us now derive how the waist of the focused Gaussian beam depends on the numerical aperture of the overfilled lens. Using the electric and magnetic field in the focal region~\cite{novotny_hecht_2012} we calculate, similar to \eqnref{eq:powerOnLens}, the power that passes through a circular aperture with radius $a$ centred at the focal point and oriented perpendicular to the $z-$axis, that is
\begin{equation}
    P(a)=2\mu_0^{-1}\int_{0}^a\text{d}\rho \rho \int_0^{2\pi}\text{d}\phi[\mathbf{E}(\rho,\phi,0)\times \mathbf{B}^*(\rho,\phi,0)]\cdot \uv_z.
\end{equation}
Analogous to a Gaussian beam we define the waist $w$ of a focused Gaussian beam as the radius of the circular aperture for which $P(w)= [1-\exp(-2)]\lim_{a\to \infty }P(a)$, that is $86\%$ of the total power is transmitted through that aperture of radius $w$. In \figref{fig2}b we show $w/\lambda_0$ for numerical apertures $\text{NA}\geq 0.1$ and a single Gaussian beam. As expected the waist decreases rapidly with increasing numerical aperture and reaches a value on the order of the wavelength $\lambda_0$ which corresponds to the diffraction limit~\cite{novotny_hecht_2012}.  

\section{Fabry-Pérot interferometer} \label{App:slab}

In this section we provide a simple model for the \LambdaName and the IRP for lenses of numerical apertures NA$\geq 0.75$ and dielectric spheres or radius $R>\lambda_0$. For these large numerical apertures the waist of the focused Gaussian beam is smaller than the wavelength, see \figref{fig2}b. It follows that the system can be modelled by a collimated beam that is normally incident on a dielectric slab of thickness $D \equiv 2R$ and relative permittivity $\epsilon$, i.e. a Fabry-Pérot interferometer. In the following we derive both the \LambdaName and the IRP for the subspace of two $x-$polarized scattering eigenmodes propagating perpendicular to the slab. These modes are given by $\mathbf{F}^\sigma_k(\rr)= A_k^\sigma f_k^p(z) \uv_x$, for $\sigma=L,R$, with normalization constant $A_k^\sigma$, eigenfrequency $\omega^\sigma_k = ck$, and
\begin{align}
    f_k^L(z) &= \exp(\im k z+\im \Phi) + r\exp(-\im k z +\im \Phi)\text{       for  }z<-D/2,\\
    f_k^L(z) &= t \exp(\im k z+\im \Phi )\text{       for  }z>D/2,\\
    f_k^R(z) &= t  \exp(-\im k z)\text{       for  }z<-D/2,\\
    f_k^R(z) &= \exp(-\im k z) + r  \exp(\im k z) \text{       for  }z>D/2.
\end{align}
Here $\Phi$ is a relative phase between the $L$ and $R$ modes. 
The reflection and transmission coefficients are given by
\begin{align}
r & = \frac{(1-\epsilon)\sin(\sqrt{\epsilon}q)}{(\epsilon+1)\sin(\sqrt{\epsilon}q)+2 \im \sqrt{\epsilon} \cos(\sqrt{\epsilon}q)},\\
t & = \frac{2\im \sqrt{\epsilon}}{(\epsilon+1)\sin(\sqrt{\epsilon}q)+2 \im \sqrt{\epsilon} \cos(\sqrt{\epsilon}q)},
\end{align}
where $q\equiv  k D$. The coupling rates $g_{\sigma \sigma '}$, analogous to the main text, are determined via the radiation pressure operator, see \eqnref{eq:definition_F}. Due to the simple geometry they can readily be shown to read
\begin{align}
    g_{LL} & \propto |r|^2  ,\\
    g_{LR} & \propto - \im |r|^2 \text{Im}(t/r)\exp(\im\Phi),\\
    g_{RL} & \propto \im |r|^2 \text{Im}(t/r)\exp(-\im\Phi),\\
    g_{RR} & \propto - |r|^2 .
\end{align}
First, let us assume the scenario where the $L$ mode is in a coherent state. In absence of the slab this corresponds to a plane wave propagating along the positive $z-$axis. It follows that the recoil heating rate reads
\begin{align}
    \Gamma \propto \sum_{\sigma}|g_{L\sigma}|^2 = |g_{LL}|^2 + |g_{LR}|^2=|r|^4\left(1+[\text{Im}(t/r)]^2\right)=g_0^2 |r|^2.
\end{align}
$\Gamma$ is maximal when the reflectance $|r|^2$ is maximal, which happens periodically at $\sqrt{\epsilon}q= \pi/2 + n \pi$, with $n \in \mathds{N}_0$. Note that these maxima also coincide with a maximal number of backscattered photons. 

Second, let us assume that both the $L$ and $R$ mode are in a coherent state. In absence of the slab this corresponds to a standing wave wave with an intensity $\cos^2(\Phi/2)$ at the origin. It follows that the recoil heating rate reads
\begin{align}
    \Gamma \propto \sum_{\sigma}|g_{L\sigma}+g_{R\sigma}|^2=|g_{LL}+g_{RL}|^2 + |g_{LR}+g_{RR}|^2.
\end{align}
Inserting the coupling rates one immediately obtains that the two terms, which represent left-oriented IRP and right-oriented IRP, respectively, are equal up to an interference term $\propto \pm \sin\Phi$, namely
\begin{align}
    |g_{LL}+g_{RL}|^2&=|r|^{2}+2|r|^4 \text{Im}(t/r) \sin\Phi,\\
    |g_{LR}+g_{RR}|^2&=|r|^{2}-2|r|^4 \text{Im}(t/r) \sin\Phi.
\end{align}
Therefore, the recoil heating rate, which is obtained by summing over the two terms, i.e. $\Gamma \propto 2|r|^2$ does not depend on $\Phi$. Furthermore, the left-oriented IRP and right-oriented IRP are only different for $\Phi \neq 0,\pi$, and they oscillate from left to right as a function of $D$.

\section{Figures} \label{app:figures}

\begin{figure}[t!]
    \centering
    \includegraphics[width=0.5\textwidth]{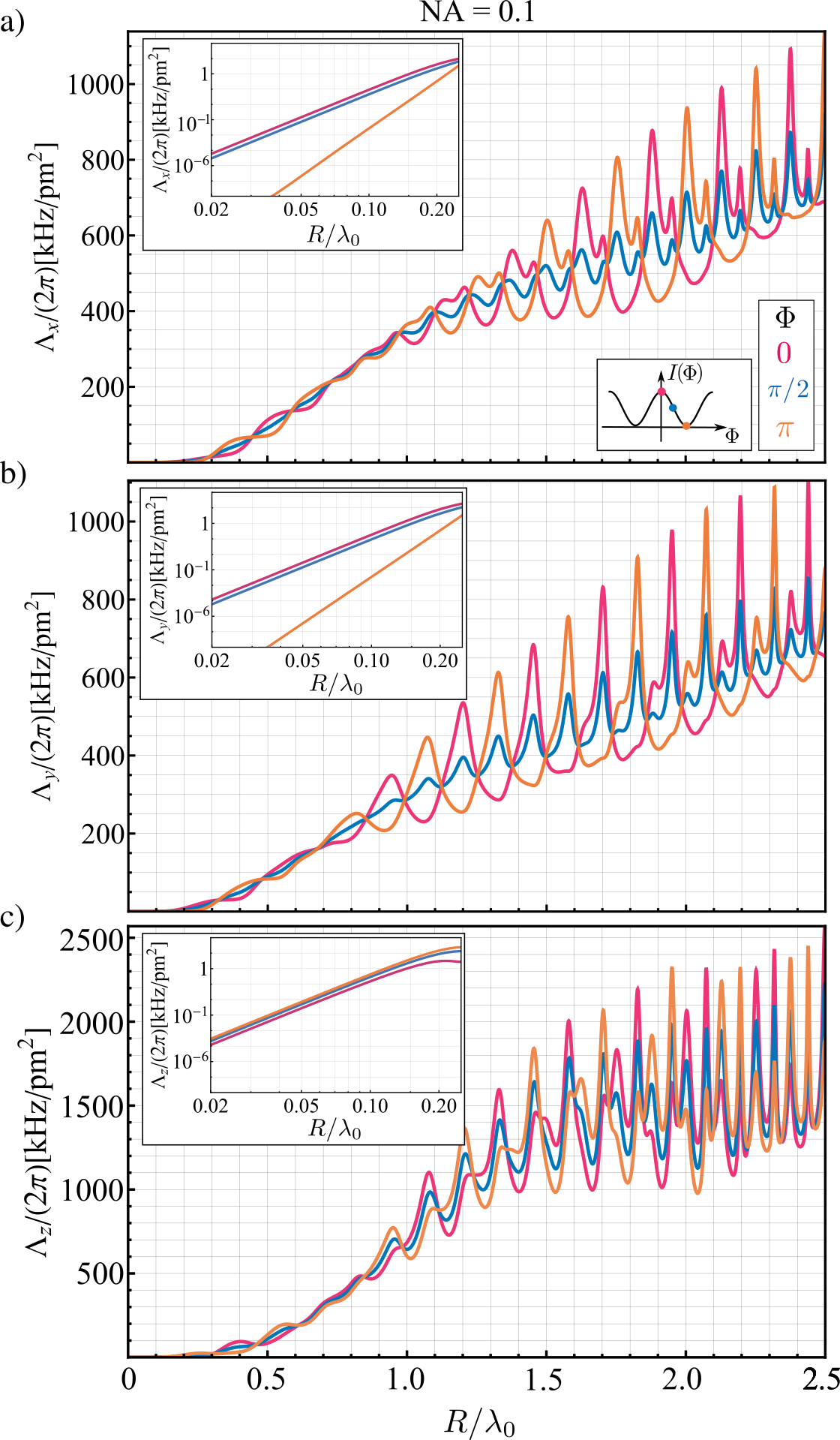}
    \caption{\LambdaNameCapitalized $\Lambda_\mu = \Gamma_\mu / r_{0\mu}^2$ for two focused $x-$polarized Gaussian beam counter-propagating along the $z-$axis as a function of the silica sphere's radius $R/\lambda_0$ and for all three axes $\mu=x,y,z$ in panel a)-c) respectively. The values for the power, wavelength, and relative permittivity are listed in Tab.~\ref{tab:parameters}.  Each panel shows the \LambdaName for $\Phi= 0, \pi/2, \pi$, $\text{NA}=0.10$, and an inset with a detailed view of the small-particle regime for all three relative phases in a log-log plot. Inset in panel a) maps the relative phase $\Phi$ to the corresponding intensity at the origin. }
    \label{fig9}
\end{figure}

\begin{figure}[t!]
    \centering
    \includegraphics[width=0.7\textwidth]{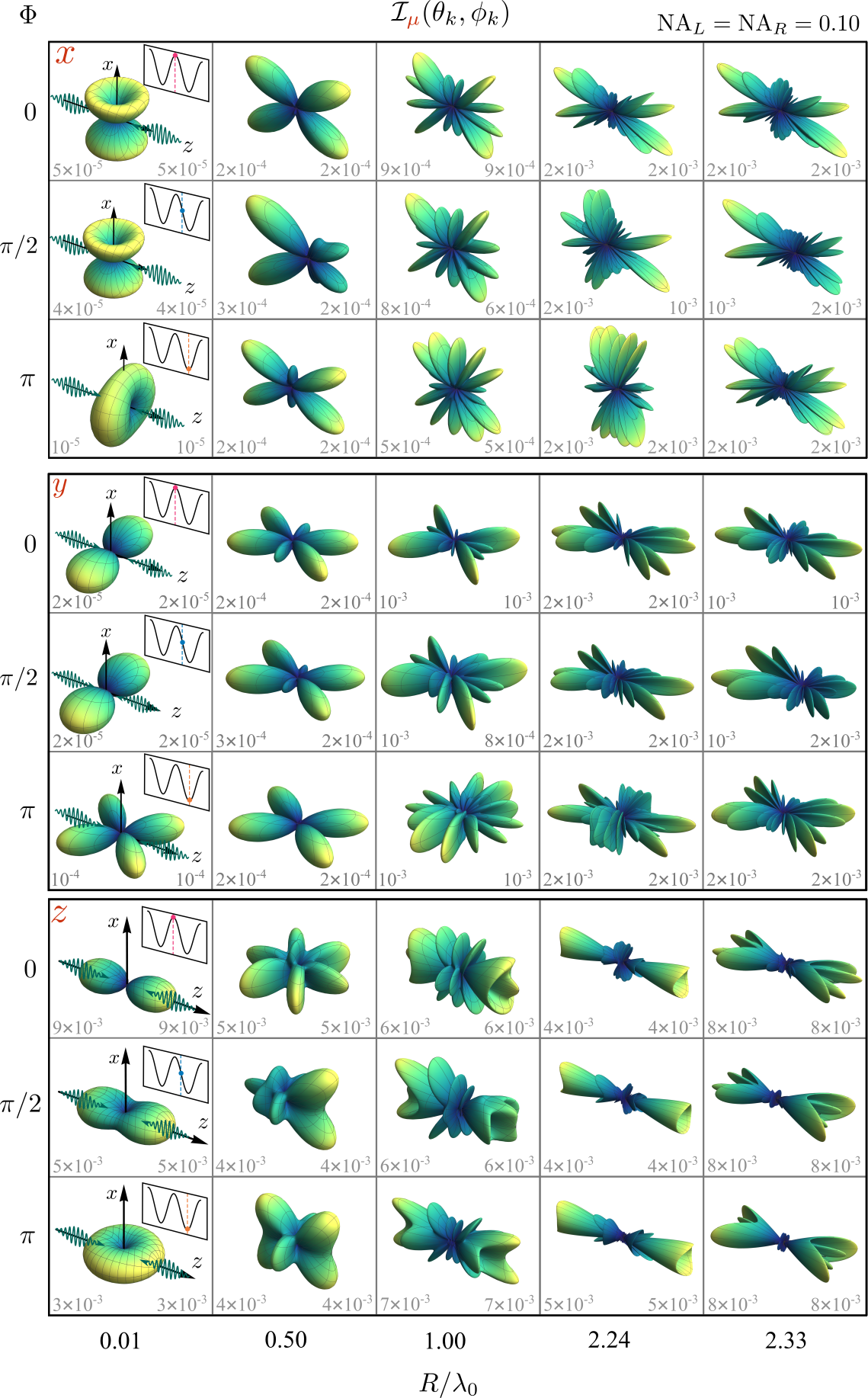}
    \caption{Information radiation patterns $\mathcal{I}_z(\theta_k,\phi_k)$ of a silica sphere, two focused $x-$polarized Gaussian beams counter-propagating parallel to the $z-$axis (reference frame in first row), and relative phases $\Phi=0,\pi/2,\pi$. The value of the IRP is encoded both in the radial distance from the center and the color scale. The two focusing lenses have a numerical aperture $\text{NA}_L=\text{NA}_R = 0.10$.  The detection efficiency for the left and right lens is shown in each sub-panel (highlighted in blue for $\eta_\mu^d > 1/9$). Across panels the value of $R/\lambda_0$ is, for each column, constant and indicated below the last row.}
    \label{fig10}
\end{figure}

\begin{figure}[t!]
    \centering
    \includegraphics[width=0.8\textwidth]{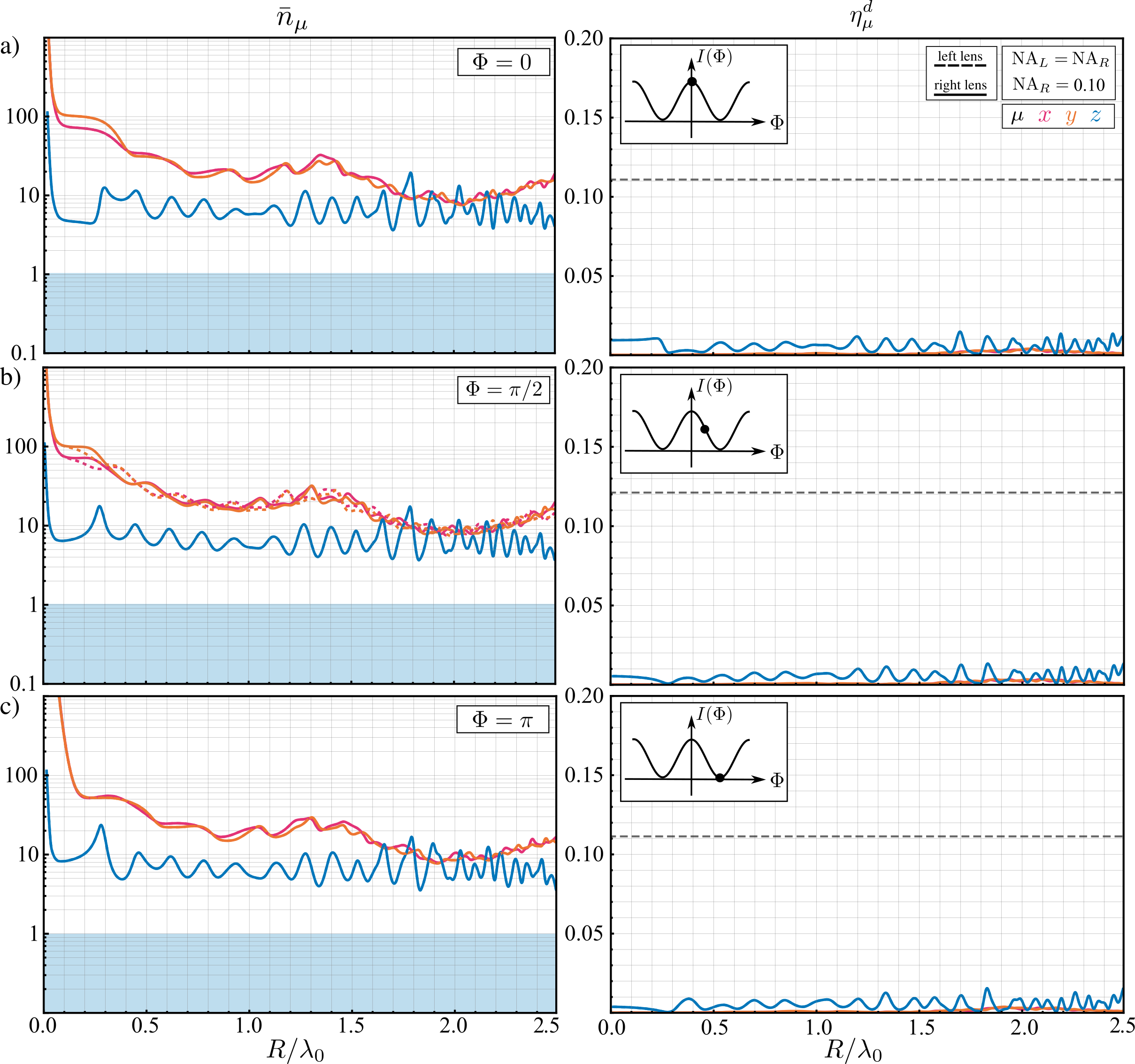}
    \caption{Detection efficiencies $\eta_\mu^d$ and minimum achievable mean phonon occupation number $\bar{n}_\mu$ along $x$, $y$, $z$ for ideal feedback as a function of $R/\lambda_0$ and $p = 10^{-9}\text{mbar}$, $T=300\text{K}$, at the same optical configuration as in \figref{fig9}, where panel a-c correspond to $\Phi = 0, \pi/2, \pi$ with an inset that maps the relative phase $\Phi$ to the corresponding intensity at the origin. The dashed (solid) lines correspond to the values at the left (right) lens. The blue shaded area highlights the region where $\bar{n}_\mu < 1$ and the grey dashed line shows $\eta_\mu^d = 1/9$ respectively.}
    \label{fig11}
\end{figure}

This section contains \figref{fig6}, \figref{fig7}, and \figref{fig8} for $\text{NA}_L=\text{NA}_R = 0.1$, shown respectively in Figures \figref{fig9}, \figref{fig10}, and  \figref{fig11}. For the low-NA standing-wave configuration the simple one-dimensional model based on the Fabry-Pérot interferometer (c.f. \appref{App:slab}) does not hold and the IRPs become again increasingly sharply peaked in the forward direction as the radius increases. In contrast to the standing-wave configuration in~\secref{sec:standingWave} the low-NA lenses lead to a much smaller detection efficiency which does not allow for ground-state cooling for almost all axes, relative phases and radii under consideration. Note however that $\bar{n}_\mu \simeq 10$ appears to be feasible for large enough spheres. 

\begin{figure}[t!]
    \centering
    \includegraphics[width=.8\textwidth]{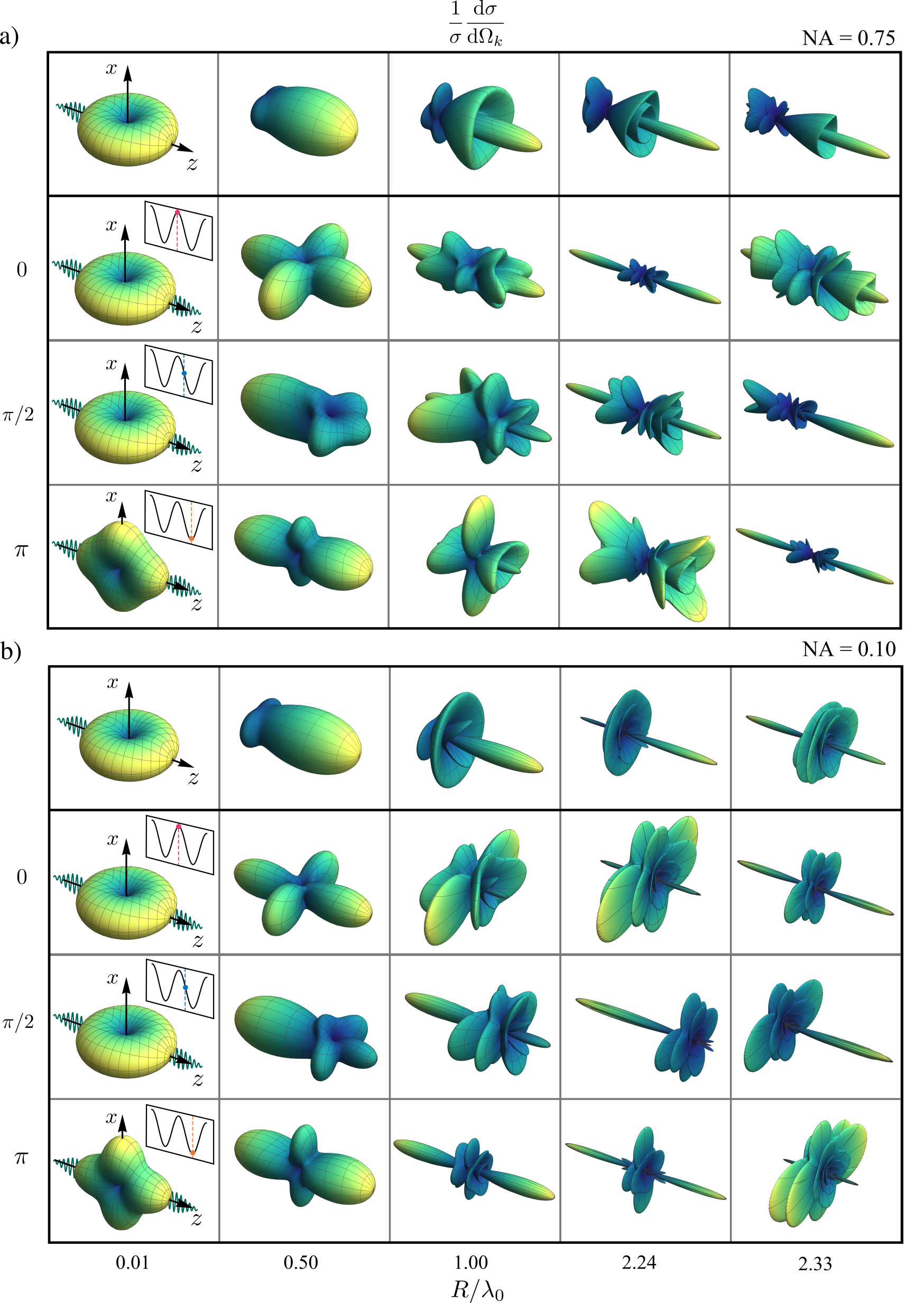}
    \caption{Normalized Lorenz-Mie differential scattering cross section $\sigma^{-1}d\sigma/d\Omega_k$ as a function of the radius for all optical configuration under consideration in the main text, i.e running- and standing-wave setup with low- and high-NA lenses, and relative phase $\Phi = 0,\pi/2,\pi$. The value of the differential scattering cross section is encoded both in the radial distance from the center and the color scale. Across panels the value of $R/\lambda_0$ is, for each column, constant and indicated below the last row.}
    \label{fig12}
\end{figure}

For completeness, we also show the normalized Lorenz-Mie differential scattering cross section $\sigma^{-1}d\sigma/d\Omega_k$~\cite{Bohren2004} for all optical configurations discussed thus in this work in \figref{fig12}.

\end{document}